\documentclass[12pt]{article}
\usepackage{amsthm,amsmath,amssymb,bbm}
\usepackage{natbib}
\usepackage{multirow}
\usepackage[pdftex]{graphicx}
\usepackage{subfigure}
\usepackage{makecell}
\usepackage{booktabs}
\usepackage{array}
\usepackage{url}
\usepackage{algorithm}
\usepackage{algorithmic}
\usepackage{bm}
\usepackage{wrapfig}
\usepackage{lipsum}
\usepackage{mathrsfs}
\usepackage{dsfont}
\usepackage{titling}
\usepackage{color}
\usepackage{multirow}
\usepackage[flushleft]{threeparttable}
\usepackage{changepage}
\usepackage{tabularx,ragged2e,booktabs,caption}
\usepackage{comment}
\usepackage[margin=1in]{geometry}
\renewcommand{\baselinestretch}{1.35}
\usepackage[colorlinks=true,allcolors=blue]{hyperref}%

\newcommand{\bcB}{\bm{\mathcal{B}}}

\newcommand{\bfeta}{{\bm\eta}}
\newcommand{\bmu}{{\bm\mu}}

\newcommand{\bGamma}{{\bm\Gamma}}

\newcommand{\bLambda}{{\bm\Lambda}}

\def\b{{\bm b}}

\def\d{{\bm d}}

\def\u{{\bm u}}

\def\w{{\bm w}}
\def\x{{\bm x}}

\def\A{{\bm A}}
\def\B{{\bm B}}

\def\D{{\bm D}}

\def\U{{\bm U}}

\newcommand\blfootnote[1]{%
  \begingroup
  \renewcommand\thefootnote{}\footnote{#1}%
  \addtocounter{footnote}{-1}%
  \endgroup
}

\begin{document}

\title{Learning Brain Connectivity in Social Cognition with Dynamic Network Regression}

\author{
Maoyu Zhang$^1$, Biao Cai$^2$, Wenlin Dai$^1$, Dehan Kong$^3$, \\
Hongyu Zhao$^2$ and Jingfei Zhang$^4$ \medskip \\
$^1$ {\normalsize Institute of Statistics and Big Data, Renmin University of China}\\ 
$^2$ {\normalsize Biostatistics Department,Yale University}\\ 
$^3$ {\normalsize Department of Statistical Sciences, University of Toronto }\\
$^4$ {\normalsize Goizueta Business School, Emory University}\\
}
\date{}
\maketitle
\blfootnote{$^*$The first two authors contributed equally to this work.}

\vspace{-0.5in}
\renewcommand{\baselinestretch}{1.15}
\begin{abstract}
Dynamic networks have been increasingly used to characterize brain connectivity that varies during resting and task states. In such characterizations, a connectivity network is typically measured at each time point for a subject over a common set of nodes representing brain regions, together with rich subject-level information. A common approach to analyzing such data is an edge-based method that models the connectivity between each pair of nodes separately. However, such approach may have limited performance when the noise level is high and the number of subjects is limited, as it does not take advantage of the inherent network structure. To better understand if and how the subject-level covariates affect the dynamic brain connectivity, we introduce a semi-parametric dynamic network response regression that relates a dynamic brain connectivity network to a vector of subject-level covariates. A key advantage of our method is to exploit the structure of dynamic imaging coefficients in the form of high-order tensors. We develop an efficient estimation algorithm and evaluate the efficacy of our approach through simulation studies. Finally, we present our results on the analysis of a task-related study on social cognition in the Human Connectome Project, where we identify known sex-specific effects on brain connectivity that cannot be inferred using alternative methods. 
\end{abstract}

\newpage
\baselineskip=21.5pt
\section{Introduction}\label{intro}

Social cognition, which refers to how individuals process, memorize, and use information in social contexts to explain and predict their own behavior and that of others \citep{fiske1991social}, is a crucial aspect of human functioning and has been extensively studied in the field of psychology and neuroscience \citep{lieberman2007social,saxe2013people}. The use of neuroimaging techniques, particularly functional magnetic resonance imaging (fMRI), has enabled a better understanding of the neural mechanisms underlying social cognition \citep{saxe2013people}. Previous studies using fMRI have shown that specific brain regions, such as the medial prefrontal cortex, the temporoparietal junction, and the superior temporal sulcus, are consistently activated during tasks related to social cognition \citep{castelli2000movement,gallagher2003functional}. 
While significant progress has been made in uncovering the neural mechanisms underlying social cognition, our understandings of the coordination between brain regions during social cognition and how it relates to individual differences in social behavior remain limited \citep{adolphs2009social}.

The social cognition study in the Human Connectome Project (HCP) \footnote{\url{https://www.humanconnectome.org/}} provided a unique opportunity for advancing our understandings of the brain connectivity underlying social cognition. In this study, imaging scans are collected using fMRI from a set of subjects as each subject goes through a sequence of cognitive tasks and rest states. In addition, it also collects subject features such as sex and social covariates (e.g., social distress). See more details in Section \ref{sec:hcp}. Based on the imaging scans, a dynamic connectivity network, characterizing activation and deactivation of connections between brain regions during task and rest states, can be constructed for each subject, with nodes corresponding to a common set of brain regions, and the edges encoding dynamic functional associations between the regions. 
From this study, it is of fundamental scientific interest to understand which brain regions are co-activated during the cognitive tasks. In addition, it is important to understand whether there are sex differences in brain connectivity during cognitive tasks, and if so, how social covariates influence these differences.

There is some recent literature on modeling a collection of networks, including dynamic networks. However, these methods may not flexibly associate dynamic network connectivity with external covariates while taking into account the structure of the network and smoothness in the dynamic brain connectivity.
Specifically, \cite{xu2014dynamic,pensky2016dynamic,ZhangCao2017,zhang2020mixed} proposed several approaches based on stochastic block models. These methods cannot associate network connectivity with external covariates. 
\cite{wang2017bayesian} proposed a Bayesian network model with covariates, which is flexible but can be computationally intensive, especially for large networks or a large number of covariates. \cite{kong2020l2rm,hu2021nonparametric,zhang2022generalized} studied matrix or network response regressions but they focused on non time-varying networks.
\cite{zhang2017tensor,hao2021sparse,zhou2021partially,tang2020individualized} considered tensor regressions that can be formulated to tackle our problem by stacking the dynamic networks observed at different time points into a tensor, but these approaches could not account for the temporal smoothness in the dynamic brain connectivity.

To model the dynamic brain connectivity in the social cognition study, we propose a new semi-parametric dynamic network model for a collection of dynamic networks with subject-level covariates. We adopt the form of generalized linear model (GLM) and assume the connectivity between a pair of regions, after a proper transformation, is the sum of two functional components. The first component is the baseline time-varying connectivity shared by all subjects and the second component involves time-varying slopes and models the effects of subject-level covariates on the time-varying brain connectivity. 
To estimate the unknown functional coefficients, we consider a nonparameteric estimation via B-spline approximations. Under such approximations, we can then write our model in the form of a dynamic network regression, where the response is the dynamic connectivity matrix and the predictors are subject covariates. With the B-spline basis, the baseline connectivity can be characterized using an intercept tensor and the covariate effect using a slope tensor. We assume the intercept tensor is low-rank and the slope tensors are structurally sparse. We discuss the benefit of placing different assumptions on these two tensor coefficients in Section \ref{sec:model}. These structural hypotheses significantly reduce the number of free parameters, facilitate model interpretability and estimability, and are commonly considered in scientific applications \citep{bi2018multilayer,zhang2022generalized}.

For estimation, we propose an efficient alternating gradient descent algorithm with a fast iterative shrinkage-thresholding method to estimate the sparse slope tensor. 
In Section \ref{sec:3}, we demonstrate in simulation studies that our method can accurately estimate the model coefficients and identify nonzero covariate effects whereas other methods fail to offer accurate estimates. 
In Section \ref{sec:4}, we apply our proposed method to the social cognition study and identify sex differences both in the baseline connectivity and social covariate effects. The majority of our results agree with the existing findings in the neuroscience literature. 
We also implement an element-wise (i.e., edge-based) method, where the results are highly noisy and lack interpretability, and a method designed for non time-varying networks \citep{zhang2022generalized}, where the results are highly sparse and cannot identify areas that are known to be engaged in social cognition. Finally, we consider a permutation based procedure to evaluate the identified sex-specific differences from our analysis.

Taken together, our work proposes a new dynamic network regression for analyzing task-evoked brain connectivity with subject-level covariates that exploits the structure in the brain network and the temporal smoothness in the time-varying connectivity. We demonstrate in simulations and real data analysis that the proposed method usually performs better than element-wise methods that model the connectivity between each pair of nodes separately. Next, we discuss in detail the motivating scientific problem and the research questions to be addressed.

\subsection{The HCP social cognition study and research questions}\label{sec:hcp}

\begin{figure}[!t]
\centering
\includegraphics[trim=0 1cm 0 0, scale=0.75]{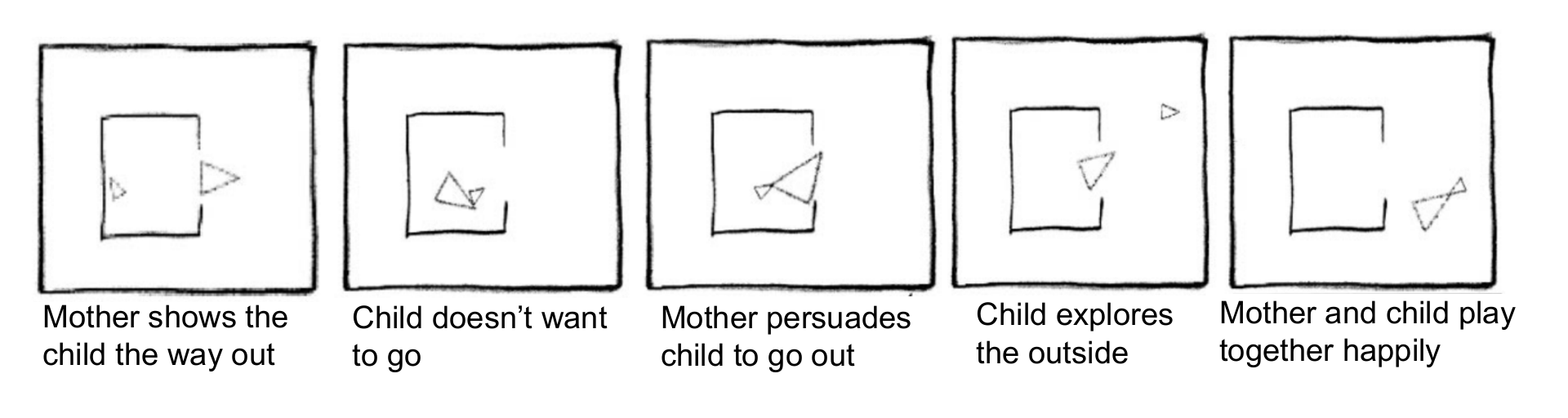}
\caption{The still illustration of a mental video. The captions, taken from \citet{castelli2000movement}, have been added for clarification and are not part of the video and are not suggested to the viewer.}\label{fig1}
\end{figure}

The social cognition study in the HCP data collected behavioral and task-related fMRI data from 850 healthy adult subjects. In each session, a participate was presented with several short videos of objects (squares, circles, triangles) interacting \citep{castelli2000movement} and the fMRI data were collected on 274 evenly spaced time points. These videos were developed by either Castelli and colleagues \citep{castelli2000movement} or Martin and colleagues \citep{wheatley2007understanding}. Specifically, two types of video clips were shown to the subjects including mental (objects interact in some way) and random (objects move randomly). Figure \ref{fig1} shows an example of the mental video block. For each participant, there were 5 video blocks (3 mental and 2 random), with each video task and rest duration taking up 23 seconds and 15 seconds, respectively. We focus our analysis on the $N=843$ subjects who were shown videos in the sequence of mental, mental, random, mental and random. Additionally, social related traits such as social distress, social support and companionship were measured for each subject via self-reported questionnaires. See more details in Section \ref{sec:4}.

In our analysis, the fMRI data are preprocessed and summarized as a $68\times 274$ spatial-temporal matrix for each subject using the Desikan-Killiany Atlas \citep{desikan2006automated} with $n = 68$ regions of interest (ROIs; see Table \ref{68region}). As each subject goes through various tasks and rest states during the scanning session and activation/deactivation of brain regions measured via fMRI are typically lagged \citep{scholvinck2010neural}, it is more appropriate to study the brain connectivity as a dynamic network. Specifically, for each subject, the dynamic network is constructed by calculating a sequence of connectivity matrices over $T$ sliding windows, each summarizing the connectivity between 68 brain regions in a given window. While there are many choices of connectivity measures \citep{smith2013functional}, the most commonly used one is perhaps the marginal Pearson correlation coefficient. We follow the vast majority of the neuroscience literature and measure connectivity in each individual by calculating Pearson correlations using samples from a pair of regions. The correlation matrix is then converted into a binary network to represent networks amongst ROIs. See more details in Section \ref{sec:4}. In our analysis, we have also considered partial correlation matrices \citep{meinshausen2006high}, and found that our main results and qualitative findings remain similar.


A number of scientifically important questions are to be addressed for this study. 
\textit{First}, which brain regions are activated during these cognitive social tasks and how do these regions function together. 
\textit{Second}, if and how subject's social covariates, such as social distress, affect the task-evoked brain connectivity. 
\textit{Third}, whether sex differences in brain connectivity during cognitive tasks exist, and if so, how do social covariates influence these differences.

We organize our paper as follows. Section \ref{sec:2} introduces the dynamic network response model and the estimation algorithm. Section \ref{sec:3} presents the simulations, and Section \ref{sec:4} analyzes the task-related study on social cognition and discusses our findings in answering the aforementioned research questions. Section \ref{sec:dis} concludes the paper with a short discussion.

\section{Model}\label{sec:2}
\subsection{Notation}
Throughout this paper, we employ the following notation. 
Let $\circ$ denote the outer product and $[k]=\{1,2\dots,k\}$.
For a vector $\b\in\mathbb{R}^{d_1}$, let $\|\b\|_2$ denote its Euclidean norm. 
For a matrix $\B\in\mathbb{R}^{d_1\times d_2}$, let $\B_{i\cdot}$ and $\B_{\cdot j}$ denote its $i$-th row and $j$-th column, respectively. 
For a tensor $\bcB \in \mathbb{R}^{d_1\times d_2\times d_3}$, let $\bcB_{ijk}$ denotes its $(i,j,k)$th entry, $\bcB_{ij\cdot}$ denote the $(i,j)$th tube fiber, and $\bcB_{\cdot\cdot k}$ denote the $k$th frontal slice. 
For $\b\in\mathbb{R}^{d_3}$ and $\bcB \in \mathbb{R}^{d_1\times d_2\times d_3}$, we define the tensor vector multiplication as 
\begin{equation}\label{eqn:tvprod}
\bcB\times_{3}\b=\sum_{k=1}^{d_3} \b_k\bcB_{\cdot\cdot k}.
\end{equation}

\subsection{The Dynamic Network Response Model}\label{sec:model}
Consider dynamic networks denoted by $\mathcal{G}_i(\mathcal{V}, \mathcal{E}_i(t))$, $i\in[N]$, observed from $N$ subjects, where $\mathcal{V} $ represents the common set of $n$ nodes and $ \mathcal{E}_i(t) $ represents the set of edges at time point $t$ for subject $i$. 
For each subject, we also observe a $p$-vector of covariates, denoted by $ \x_i=(x_{i1},\ldots, x_{ip})^T$.  
At each time point $ t $, the network $\mathcal{G}_i(\mathcal{V}, \mathcal{E}_i(t))$ can be uniquely represented by its $ n\times n $ adjacency matrix $ \A^{(i)}(t) $, where $ \A^{(i)}_{jj'}(t)$ denotes the edge between nodes $ j $ and $ j' $ at time point $t$ in subject $ i $. 
The edges can be 
continuous, binary or nonnegative integers.
Without loss of generality, we assume $ t\in [0,1] $, and $\A^{(i)}(t)$ are observed at $T$ 
time points $\{t_1,t_2,\ldots,t_T\}$ such that $ 0= t_1\leq t_2\leq \ldots \leq t_T=1 $. 

Let $\bmu^{(i)}(t)=\mathbb{E}(\A^{(i)}(t)|\x_i)$, where the expectation $\mathbb{E}(\cdot)$ is applied element-wise to entries in $\A^{(i)}(t)$.
We assume that, conditioning on $\x_i$, the entries in $\A^{(i)}(t)$ are independent and follow an exponential distribution with a canonical link function that
\begin{equation}\label{model2}
g(\bmu^{(i)}(t))=\B_0(t)+\sum_{l=1}^p x_{il}\B_l(t),\quad i=1,\ldots,N,
\end{equation}
where $\B_0(t)\in\mathbb{R}^{n\times n}$ characterizes the population-level time-varying network connectivity and $\B_l(t)\in\mathbb{R}^{n\times n}$ characterizes the time-varying effects of the $l$-th covariate on the network connectivity. 
The function $g(\cdot)$ is an invertible link function, as commonly used in GLMs \citep{mccullagh1989generalized}, and is applied element-wise to entries in $\bmu^{(i)}(t)$.

\begin{figure}[t!]
\centering
\includegraphics[trim=0 1cm 0 0, width=\linewidth]{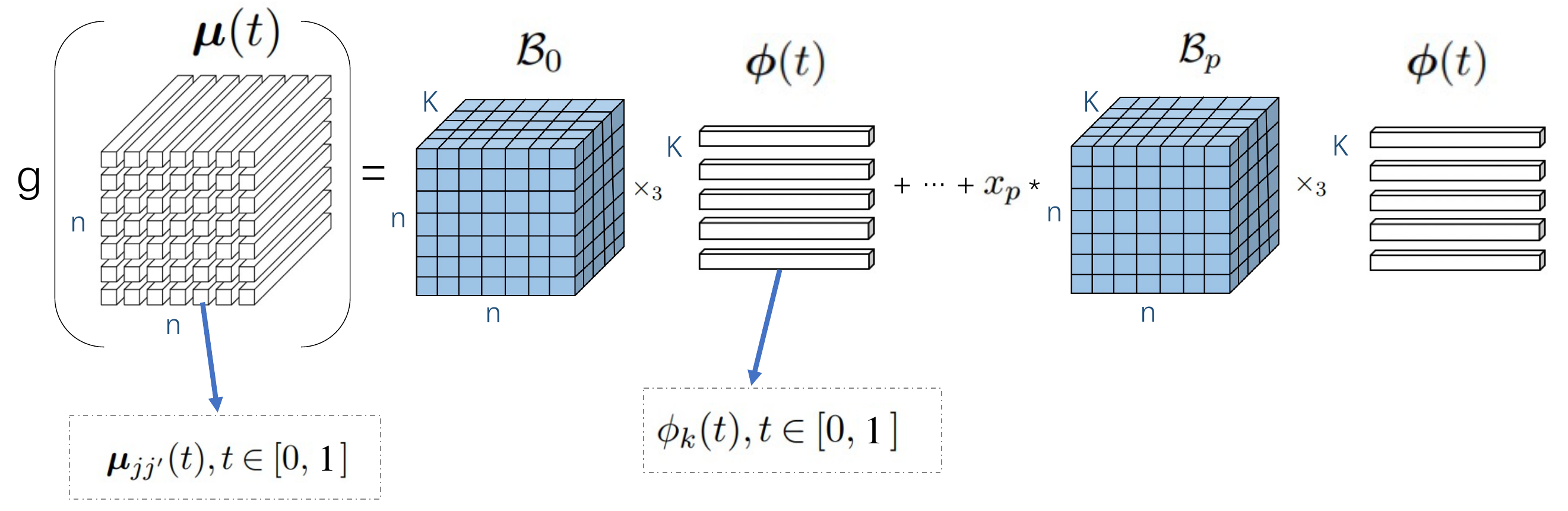}
\caption{An illustration of the dynamic network response model.} \label{Fig:MODEL}
\end{figure}

Let $\B_{ljj'}(t)$ denote the $(j,j')$th element of $\B_l(t)$. 
To estimate the unknown functions $\B_{ljj'}(t)$'s, we consider a nonparametric estimation using B-spline approximations. 
Specifically, we approximate $\B_{ljj'}(t)$'s using a $K$-dimensional basis denoted by $ {\boldsymbol \phi}(t)=(\phi_1(t),\ldots, \phi_K(t))^T $ such that 
$ \B_{ljj'}(t)={\boldsymbol \phi}^T(t)\times\b_{ljj'}+r_{ljj'}(t)$, where $\b_{ljj'}\in\mathbb{R}^K$ and $r_{ljj'}(\cdot)$ is the approximation residual. 
Defining $ \mathcal{B}_l\in\mathbb{R}^{n\times n\times K}$ such that $\mathcal{B}_{ljj'}= \b_{ljj'}$ for all $j,j'$ and $l$, model (\ref{model2}) can be rewritten as
\begin{equation}\label{approximatemodel}
g(\bmu^{(i)}(t))=\mathcal{B}_0\times_3 {\boldsymbol \phi}(t)+\sum_{l=1}^p x_{il}(\mathcal{B}_l\times_3 {\boldsymbol \phi}(t)), 
\end{equation}
where $\times_3$ is defined as in \eqref{eqn:tvprod}, $\mathcal{B}_0,\ldots,\mathcal{B}_p$ are unknown tensor coefficients of dimension $n\times n\times K$. A graphical illustration of model \eqref{approximatemodel} is given in Figure \ref{Fig:MODEL}. 

One challenge in estimating model \eqref{approximatemodel} is the inherent high-dimensionality of the tensor coefficients. In our analysis of the HCP social cognition study, each coefficient tensor $\mathcal{B}_l$ is of dimension $68\times 68\times 10=46,240$, far exceeding the number of subjects in the study. Thus, it is imperative to employ effective dimension reduction assumptions that can facilitate estimability and interpretability. Next, we move to discuss the dimension reduction assumptions placed on the baseline effect coefficient tensor $\mathcal{B}_0$ and the covariate effect coefficient tensors $\mathcal{B}_1,\ldots,\mathcal{B}_p$. We also discuss the need for considering different assumptions for these two types of effects. 

\smallskip\noindent
\textbf{Low-rankness on $\mathcal{B}_0$.}
The component $\mathcal{B}_0$ is the baseline coefficient tensor and we assume that it possesses a low-rank structure. 
This specification assumes that there is a low-dimensional structure in the baseline time-varying network connectivity, such that both the nodes and the basis coefficients have lower dimensional representations. 
This is similar to, but more general than, for example, the stochastic blockmodel \citep{holland1983stochastic}, a well-studied network model that assumes the nodes form a number of groups and after reorganizing by group membership, the connecting probability matrix is a block matrix. 

In our data problem, the low-rank assumption effectively reduces the number of parameters and increases computational efficiency.
Specifically, we assume that $\mathcal{B}_0$ admits the following rank-$R$ CP decomposition \citep{kolda2009tensor}: 
$$
\mathcal{B}_0=\sum_{r=1}^R w_{r}\u_{1r} \circ \u_{1r} \circ \u_{3r},
$$
where $w_r\in\mathbb{R}^+$, $\u_{1r}\in\mathbb{R}^n$ and $\u_{3r}\in\mathbb{R}^K$. 
For identifiability, we assume $\u_{1r}$'s and $\u_{3r}$'s are unit length vectors.  
We note that the above formulation is for undirected networks.
When the networks are directed, we can write $\mathcal{B}_0=\sum_{r=1}^R w_{r}\u_{1r} \circ \u_{2r} \circ \u_{3r}$, where $\u_{2r}\in\mathbb{R}^n$ is a unit length vector.

\smallskip\noindent
\textbf{Structured sparsity in $\mathcal{B}_1,\ldots, \mathcal{B}_p$.} We assume that the subject covariates have sparse effects on the dynamic network connectivity, that is, the effects concentrate on a small number of regions. 
This is scientifically plausible, as brain connections are energy consuming and biological units tend to minimize energy-consuming activities \citep{bullmore2009complex}. Sparsity also greatly reduces the number of free parameters and improves interpretation of the resulting model.
Specifically, we assume that $ \mathcal{B}_l$, $l\in[p]$, is structurally sparse in that it has sparse nonzero tube fibers, corresponding to sparse nonzero time-varying effects $\B_{ljj'}(t)$, $l\in[p]$. 
To encourage structural sparsity, we consider the group lasso \citep{yuan2006model} penalty, defined as
\begin{equation}\label{eqn:gl}
\mathcal{P}(\mathcal{B}_1,\ldots,\mathcal{B}_p ) = \sum_{l=1}^{p}\sum_{j\neq j'}^n\|\mathcal{B}_{ljj'\cdot}\|_2.
\end{equation}

\smallskip\noindent
\textbf{Different assumptions on $\mathcal{B}_0$ and $\mathcal{B}_1,\ldots, \mathcal{B}_p$.} We briefly discuss the benefits and necessity of imposing separate structures on $\mathcal{B}_0$ and $\mathcal{B}_1,\ldots, \mathcal{B}_p$. 
It is natural to think that one could stack $\mathcal{B}_0, \mathcal{B}_1,\ldots, \mathcal{B}_p$ into one higher-order coefficient tensor of size $n\times n\times K\times (p+1)$, and specify it to be both low-rank and sparse. However, assuming $\mathcal{B}_0$ to be sparse may not be plausible in the GLM setting. 
For instance, when the network edges are binary and $g(\cdot)$ is the logit link, $g(0)$ yields a connecting probability of 0.5; when the network edges are counts and $g(\cdot)$ is the log link, $g(0)$ is not well defined. Correspondingly, a sparse $\mathcal{B}_0$ does not necessarily imply sparsity in the baseline connectivity, and may not even be well defined. This issue is unique in using sparse GLM to model edges in a network. 
Finally, more complex structures on $\mathcal{B}_1,\ldots, \mathcal{B}_p$ can be incorporated (for example, $\mathcal{B}_1,\ldots, \mathcal{B}_p$ are low-rank and sparse), which can further reduce the number of effective parameters. However, such assumptions are expected to incur a much higher computational cost and also involve more tuning parameters on, for example, the rank of each coefficient. To balance model complexity and feasibility, we focus on the current assumption that assumes $\mathcal{B}_1,\ldots, \mathcal{B}_p$ have structured sparsity.

\subsection{Estimation}
Recall that $\mathcal{B}_0=\sum_{r=1}^R w_{r}\u_{1r} \circ \u_{1r} \circ \u_{3r}$. Denote $\w=(w_1,\ldots,w_R)$, $\U_{1}=(\u_{11},\ldots,\u_{1R})\in\mathbb{R}^{n\times R}$, $\U_{3}=(\u_{31},\ldots,\u_{3R})\in\mathbb{R}^{K\times R}$ and $\bGamma=(\mathcal{B}_1,\ldots,\mathcal{B}_p)\in\mathbb{R}^{n\times n\times K\times p}$.
Under model \eqref{approximatemodel}, the negative loglikelihood function, up to a constant, can be written as
\begin{equation}
\label{loglikelihood2}
\ell(\w,\U_1,\U_3,\bGamma)=-\frac{1}{N}\sum^N_{i=1}\sum^n_{j< j'}\sum^T_{h=1}\left[\A^{(i)}_{jj'}(t_h)\bfeta^{(i)}_{jj'}(t_h)-\psi\left\{\bfeta^{(i)}_{jj'}(t_h)\right\}\right],
\end{equation}
where $\bm\eta^{(i)}(t)=\mathcal{B}_0\times_3 {\boldsymbol \phi}(t)+\sum_{l=1}^p x_{il}(\mathcal{B}_l\times_3 {\boldsymbol \phi}(t))$, and $\psi(\cdot)$ is the cumulant function with its first derivative $\psi'(\cdot) = g(\cdot)^{-1}$ \citep{mccullagh1989generalized}.
We estimate the parameters $\w,\U_1,\U_3,$ and $\bGamma$ by solving the following optimization problem,
\begin{equation}
\label{obj1}
\min_{\w,\,\U_1,\,\U_3,\,\bGamma}\ell(\w,\U_1,\U_3,\bGamma)+\lambda \mathcal{P}(\mathcal{B}_1,\ldots,\mathcal{B}_p ),
\end{equation}
where $\mathcal{P}(\cdot)$ is as defined in \eqref{eqn:gl} and $\lambda$ is a tuning parameter.

The optimization problem in \eqref{obj1} is computationally challenging, as the size of the networks, the dimension of the covariates and the number of basis functions can be large in practice. The GLM loss function further increases the computation burden due to its nonlinearity. While \eqref{obj1} is nonconvex, the conditional optimization with respect to $\u_{1r}$, while fixing all other parameters, is convex, and the same holds for $\w$, $\u_{r3}$'s and $\mathcal{B}_j$'s. This observation permits an alternating minimization algorithm. One potential issue in such an approach is that solving for $\bGamma$, conditional on all other parameters, is a regularized optimization problem of dimension $n\times n\times K\times p$. This can be computationally expensive when the network size $n$, the number of splines $K$ and the dimension of the covariates $p$ are large. To tackle this challenge, we consider a proximal gradient descent algorithm that is easy to implement and computationally efficient. Our estimation procedure is summarized in Algorithm \ref{algo3}. 

\begin{algorithm}[!t]
\caption{Optimization procedure of \eqref{obj1}}
\begin{algorithmic}\normalsize{
\STATE \textbf{Input}: 
rank $R$, tuning parameter $\lambda$ and step size $\eta$.
\STATE\hspace{0.15in} \textit{Step 1}: initialize $\w^{(0)},\U_1^{(0)},\U_3^{(0)},\mathcal{B}^{(0)}_1,\ldots,\mathcal{B}^{(0)}_p$. 
\STATE\hspace{0.15in} {\bf Repeat} Steps 2-5 for $t=0,1,\ldots$ until convergence.
\STATE\hspace{0.25in} \textit{Step 2}: {\bf repeat} the following steps for $r=1,2,...R.$\\
\hspace{0.7in} 
$\tilde\u_{1r}^{(t+1)}=\arg\min_{\u}\ell(\w^{(t)},\u_{11}^{(t+1)},\dots,\u_{1(r-1)}^{(t+1)},\u, \dots, \u_{1R}^{(t)},\u_{31}^{(t)}, \dots, \u_{3R}^{(t)},\bGamma^{(t)})$,\\
\hspace{0.7in} $\tilde\u_{3r}^{(t+1)}=\arg\min_\u\ell(\w^{(t)},\u_{11}^{(t+1)}, \dots, \u_{1R}^{(t+1)}, \u_{31}^{(t+1)}, \dots, \u_{3(r-1)}^{(t)},\u, \dots , \u_{3R}^{(t)},\bGamma^{(t)})$.\\
\STATE\hspace{0.25in} \textit{Step 3}: $\tilde\U_{j}^{(t+1)}=(\tilde\u_{j1}^{(t+1)},\ldots,\tilde\u_{jR}^{(t+1)})$,\,\,$j=1,3$,\\
\hspace{0.7in} $\w^{(t+1)}=\w^{(t)}\text{Norm}(\tilde\U_1^{(t+1)})^2\text{Norm}(\tilde\U_3^{(t+1)})$,\\ 
\hspace{0.7in} $\U^{(t+1)}_j=\text{Unit}(\tilde\U_j^{(t+1)})$,\,\,$j=1,3$.\\
\STATE\hspace{0.25in} \textit{Step 4}: set $\bGamma^{(t,0)}=\bGamma^{(t)}$, $\bLambda^{(t,0)}=\bGamma^{(t)}$, $h_0=1$.
\STATE\hspace{0.25in} \textit{Step 5}: {\bf repeat} the following steps for $s=0,1,\ldots$ until convergence.
\STATE 
\hspace{0.7in} $\bGamma^{(t,s+1)}=\mathcal{S}_{\lambda\eta}(\bLambda^{(t,s)}-\eta\nabla_\bLambda\ell(\w^{(t+1)},\U^{(t+1)}_1,
\U^{(t+1)}_3,\bLambda)\mid_{\bLambda=\bLambda^{(t,s)}})$,\\
\hspace{0.7in} $h_{s+1}=(1+\sqrt{1+4h_s^2})/2$,\\
\hspace{0.7in}
$\bLambda^{(t,s+1)}=\bGamma^{(t,s+1)}+\frac{h_s-1}{h_{s+1}}(\bGamma^{(t,s+1)}-\bGamma^{(t,s)})$. 
\STATE\hspace{0.25in} \textit{Step 6}: set $\bGamma^{(t+1)}=\bGamma^{(t,s)}$.
\STATE \textbf{Output}: $\hat{\w},\hat\U_1,\hat\U_3,\hat\bGamma$.}
\end{algorithmic}
\label{algo3}
\end{algorithm}

In Step 2, $\tilde\u_{jr}$'s are solved using a Newton-type algorithm \citep{schnabel1985modular} and the gradients are given in Section \ref{sec:gr} in the supplement. In Step 3, we define two matrix operators for $\U$. Norm$(\U)$ calculates the $\ell_2$ norms of columns in a matrix $\U$ and Unit$(\U)$ rescales the columns of a matrix into unit vectors. That is,
\begin{equation*}
\text{Norm}(\U) = \left[\|\U_{.1}\|_2, \|\U_{.2}\|_2, \ldots, \|\U_{.R}\|_2\right]^T \text{ and } \text{Unit}(\U) = \left[\frac{\U_{.1}}{\|\U_{.1}\|_2}, \frac{\U_{.2}}{\|\U_{.2}\|_2}, \ldots, \frac{\U_{.R}}{\|\U_{.R}\|_2}\right].
\end{equation*}
In Step 5, we employ the fast iterative shrinkage-thresholding method \citep[FISTA,][]{beck2009fast} under group lasso penalty. Specifically, we define the shrinkage operator by $\mathcal{S}_{\lambda\eta}(\bGamma)=(\mathcal{T}_{\lambda\eta}(\mathcal{B}_1),\ldots,\mathcal{T}_{\lambda\eta}(\mathcal{B}_p))\in\mathbb{R}^{n\times n\times K\times p}$, where
$$
\mathcal{T}_{\lambda\eta}(\mathcal{B}_{l})_{jj'\cdot}=\left(1-\frac{\lambda\eta}{\|\mathcal{B}_{ljj'\cdot}\|_2}\right)_{+}\mathcal{B}_{ljj'\cdot},
$$
and $(x)_+=\max(0,x)$. 
In the FISTA algorithm and at step $s+1$, the iterative shrinkage
operator $\mathcal{S}_{\lambda\eta}(\cdot)$ is not directly applied to the previous point $\bGamma^{(t,s)}$, but rather at the point $\bLambda^{(t,s)}$ which
uses a specific linear combination of the previous two points $\bGamma^{(t,s)}$ and $\bGamma^{(t,s-1)}$. The FISTA algorithm has been shown to enjoy a fast global rate of convergence \citep{beck2009fast} and is easy to implement. 
The stepsize $\eta$ is typically chosen as the Lipschitz constant of $\nabla_\bGamma\ell(\w,\U_1,\U_3,\bGamma)$, which can be approximately calculated given the initial values.

\smallskip\noindent
\textbf{Initialization.} 
In Algorithm \ref{algo3}, we need to determine the initial values for the alternating minimization procedure. To obtain a good initial estimate, we first estimate $\mathcal{ B}^{(0)}_0, \mathcal{B}^{(0)}_1,\dots,\mathcal{B}^{(0)}_p$ via an element-wise generalized spline regression; see \eqref{eqn:initial}. We then estimate $\w^{(0)}, \U_1^{(0)}, \U_3^{(0)}$ via a $\mathrm{CP}$ decomposition of the estimated $\mathcal{ B}^{(0)}_0$. In our experiments, this initialization procedure leads to a good numerical performance of Algorithm \ref{algo3}. The accuracy of this initialization procedure is evaluated in Section \ref{sec:3}. 


\smallskip\noindent
\textbf{Parameter tuning.} The rank $R$ and regularization parameter $\lambda$ are two tuning parameters in our algorithm. 
We choose these parameters using the eBIC criterion that was first developed for variable selection in the diverging dimension regime in \cite{chen2012extended}. 
It has been demonstrated that 
the eBIC function is effective as a heuristic criterion to balance model fitting and complexity when used in low-rank estimation problems \citep{srivastava2017expandable,cai2021jointly,zhang2022generalized}.
Specifically, we choose the combination of $(R, \lambda)$ that minimizes,
$$N\times\ell(\hat\w,\hat\U_1,\hat\U_3,\hat\bGamma)+[\log\left(n^{2} NT/2\right)+ \log\left(n^{2} K(p+1)/2\right)]\times [R(n+K)+\sum_{l=1}^p||\hat{\mathcal{B}_l}||_0/2],$$
where $\ell$ is the loss function in (\ref{loglikelihood2}), and $\hat\w,\hat\U_1,\hat\U_3,\hat\bGamma$ are the estimates of $\w,\U_1,\U_3,\bGamma$ under the working rank and regularization parameter.
In our numerical experiments, the above eBIC is found to be minimized at the true rank and sparsity level under the selected $\lambda$.

\section{Simulation}\label{sec:3}
We conduct simulations to investigate the performance of our proposed method. We focus on symmetric networks, and compare our proposed dynamic network response regression method, referred as $\rm DNetReg$, with two alternative element-wise approaches. 

The first element-wise approach, referred as $\rm EdgeReg$, fits element-wise GLMs at each time point $t_k$. That is, for any $j,j'\in [n], h\in [T]$, consider
\begin{equation}\label{eqn:initial0}
g(\bmu^{(i)}_{j j'}(t_h))=\B_{0j j'}(t_h)+\sum_{l=1}^p x_{il}\B_{lj j'}(t_h),\quad i\in [N].
\end{equation}
This element-wise approach ignores both the network structure and the temporal smoothness in the dynamic brain connectivity. 
The second element-wise approach, referred as $\rm DEdgeReg$, fits a generalized spline regression to each entry in $\boldsymbol{A}_{j j'}(t)$. Specifically, for any $j,j'\in [n]$, consider
\begin{equation}\label{eqn:initial}
g(\bmu^{(i)}_{j j^{\prime}}(t))=\mathcal{B}_{0j j'\cdot}^\top{\boldsymbol \phi}(t)+\sum_{l=1}^p x_{il}\mathcal{B}_{lj j'\cdot}^\top\boldsymbol \phi(t), \quad i\in [N].
\end{equation}
A Newton-type algorithm is employed to estimate the parameters in the above model. 
The method $\rm DEdgeReg$ is used to find the initial values in Algorithm \ref{algo3}.


\begin{table}[!t]
\small
\centering
\setlength{\tabcolsep}{2pt}
\caption{Simulation results under the generalized dynamic network response model with $N=50$ and varying numbers of nodes $n$, rank $R$ and sparsity proportion $s_0$. Marked in boldface are those achieving the best evaluation criteria in each setting.}\label{result_sim}
\begin{tabular}{|c|c|c|llllll|}
\hline $n$ & $R$ & $s_0$ & Method & Error of $\boldsymbol{\mu}^{(i)}(t)$ & Error of $\mathcal{B}_0$ & Error of $\mathcal{B}_1$ &TPR &FPR  \\\hline
 \multirow{12}{*}{50} & \multirow{6}{*}{2}         & \multirow{3}{*}{0.05}   &$\rm EdgeReg$ & 31.986(0.759)  & -  &  -   &   0.010(0.051) &0        \\
                            &   &   & $\rm DEdgeReg$ & 8.767(0.850)  & 25.010(9.601)  &  14.599(1.256)   &  -  &-      \\
                            &   &   & $\rm DNetReg$ & \bf  2.410(0.306)  & \bf 5.925(1.048)  &  \bf 7.054(0.727)   &   \bf 1.000(0.000) &   0.016(0.019)       \\
                       \cline{3-9} 
                                               &                                              & \multirow{3}{*}{0.1}   &$\rm EdgeReg$ & 31.912(0.724)  & -  &  -   &   0.012(0.072) &0        \\
                            &   &   & $\rm DEdgeReg$ & 8.636(0.235)  & 25.588(7.593)  &  17.394(1.279)   &  -  &-      \\
                            &   &   &  $\rm DNetReg$ & \bf 3.067(0.448)  & \bf 6.545(1.026)  &  \bf 9.774(0.886)   &   \bf 1.000(0.000) &  0.017(0.016)       \\
                       \cline{2-9} 
                                              & \multirow{6}{*}{5}                  & \multirow{3}{*}{0.05} & $\rm EdgeReg$  &  29.921(0.718)  & -  &  -   &  0.001( 0.005) &0        \\
                             &   &   & $\rm DEdgeReg$ & 8.225(0.213)  & 35.912(11.527) &  16.348(1.328)  &  -   &-    \\
                            &   &   &  $\rm DNetReg$ & \bf 2.875(0.203)  & \bf 7.896(0.935)  & \bf7.546(0.791)   & \bf 1.000(0.000)& 0.020(0.025)       \\
                       \cline{3-9} 
                                               &                                              & \multirow{3}{*}{0.1} & $\rm EdgeReg$  &  29.878(0.799)  & -  &  -   &  0.006(0.031) &0        \\
                             &   &   & $\rm DEdgeReg$ & 8.340(0.213)  & 36.304(11.432) &  18.528(1.652)   &  -   &-    \\
                            &   &   &  $\rm DNetReg$ & \bf 3.436(0.146)  & \bf 8.428(1.114)  & \bf 10.833(1.317)   & \bf 1.000(0.000)& 0.021(0.021)       \\
\cline{1-9} 
  \multirow{12}{*}{100}  & \multirow{6}{*}{2} &  \multirow{3}{*}{0.05}   & $\rm EdgeReg$ & 64.302(1.125)  & -  &  -   & 0.000(0.000) &0        \\
    &   &   & $\rm DEdgeReg$ & 17.461(0.532)  & 52.495(18.720)  & 28.717(1.847)   &  -    &-    \\
    &   &   &  $\rm DNetReg$ & \bf 4.556(0.371)  & \bf 10.441(1.991)  &  \bf 14.095(1.289)   &  \bf 1.000(0.000) & 0.016(0.014)  
\\
\cline{3-9} 
                           &   & \multirow{3}{*}{0.1}   & $\rm EdgeReg$ & 64.170(1.081)  & -  &  -   & 0.000(0.000) &0        \\
    &   &   & $\rm DEdgeReg$ & 17.372(0.396)  & 50.158(10.227)  & 31.699(1.893)   &  -    &-    \\
    &   &   &  $\rm DNetReg$ & \bf 5.617(0.295)  & \bf 10.844(1.658)  &  \bf 19.895(1.818)   &  \bf 1.000(0.000) & 0.015(0.014)  
\\
                       \cline{2-9} 
                                               & \multirow{6}{*}{5}                  & \multirow{3}{*}{0.05}   &$\rm EdgeReg$ &  59.413(1.667)  & -  &  -   & 0.000(0.000) &0         \\
                       
                           &   &   & $\rm DEdgeReg$ & 16.491(0.353)  & 68.531(12.035)  &  32.981(1.898)   &  -     &-   \\
                       
                           &   &   &  $\rm DNetReg$ & \bf 5.359(0.435)  & \bf 11.945(2.530)  &  \bf 15.242(1.551)   &  \bf 1.000(0.000) &0.020(0.019)        \\
                       \cline{3-9} 
                                               &                                              & \multirow{3}{*}{0.1}   &$\rm EdgeReg$ &  59.554(1.463)  & -  &  -   & 0.000(0.000) &0         \\
                       
                           &   &   & $\rm DEdgeReg$ & 16.978(2.029)  & 68.683(11.474)  &  34.618(2.946)   &  -     &-   \\
                       
                           &   &   &  $\rm DNetReg$ & \bf 6.418(0.472)  & \bf 12.361(2.131)  &  \bf 21.451(1.961)   &  \bf 1.000(0.000) &0.019(0.015)        \\
                       \hline
   \end{tabular}
\end{table}

\begin{table}[!t]
\small
\centering
\setlength{\tabcolsep}{2pt}
\caption{Simulation results under the generalized dynamic network response model with $N=100$ and varying numbers of nodes $n$, rank $R$ and sparsity proportion $s_0$. Marked in boldface are those achieving the best evaluation criteria in each setting.}\label{result_sim_100}
\begin{tabular}{|c|c|c|llllll|}
\hline $n$ & $R$ & $s_0$ & Method & Error of $\boldsymbol{\mu}^{(i)}(t)$ & Error of $\mathcal{B}_0$ & Error of $\mathcal{B}_1$ &TPR &FPR  \\\hline
 \multirow{12}{*}{50} & \multirow{6}{*}{2}         & \multirow{3}{*}{0.05}   &$\rm EdgeReg$  & 31.976 (0.783)  & -  &   -  &  0.010 (0.054)&     0   \\
                            &   &   & $\rm DEdgeReg$  & 8.695 (0.202)  & 23.381 (2.607)  & 14.344 (0.898)    & - &  -      \\
                            &   &   & $\rm DNetReg$  & \bf 1.833(0.202)  &\bf 4.150 (1.022)   &  \bf  4.772(0.384)   & \bf 1.000 (0.000)  &   0.017(0.018)     \\
                       \cline{3-9} 
                                               &    & \multirow{3}{*}{0.1}   &$\rm EdgeReg$  & 31.928 (0.746)  &-   & -    & 0.012 (0.073) & 0       \\
                            &   &   & $\rm DEdgeReg$  & 8.680 (0.200)  & 23.153 (2.358)  & 14.835 (0.873)    & - & -       \\
                            &   &   &  $\rm DNetReg$ &  \bf 2.256 (0.217) & \bf 5.010 (1.505)  & \bf 6.796 (0.497)    &  \bf 1.000 (0.000)  &  0.018(0.014)        \\
                       \cline{2-9} 
                                              & \multirow{6}{*}{5} & \multirow{3}{*}{0.05} & $\rm EdgeReg$  &  29.982 (0.725)  & -  &  -   & 0.006 (0.038)&0        \\
                             &   &   & $\rm DEdgeReg$ &  8.243 (0.204) & 33.525 (6.680)  & 16.416 (1.254)    & -& -       \\
                            &   &   &  $\rm DNetReg$  & \bf 2.369 (0.277)  &\bf 6.896 (0.823)   &\bf 5.437 (0.585)     & \bf 1.000 (0.000) & 0.014 (0.015)       \\
                       \cline{3-9} 
                                               &     & \multirow{3}{*}{0.1} & $\rm EdgeReg$  & 29.888 (0.745)  & -  &  -   & 0.002 (0.004) &0        \\
                             &   &   & $\rm DEdgeReg$  & 8.240 (0.194)  & 34.872 (9.490)  & 17.173 (1.386)    &-  & -       \\
                            &   &   &  $\rm DNetReg$  & \bf 2.605 (0.162)  & \bf 7.068 (1.163)   &  \bf 7.722 (0.794)   & \bf 1.000 (0.000)  & 0.013 (0.014)       \\
\cline{1-9} 
  \multirow{12}{*}{100}  & \multirow{6}{*}{2} &  \multirow{3}{*}{0.05}   & $\rm EdgeReg$  & 64.305(1.136)  & -  & -    &  0& 0       \\
    &   &   & $\rm DEdgeReg$  &  17.428 (0.384) & 48.244 (5.731)  & 28.645 (1.863)     & - & -       \\
    &   &   &  $\rm DNetReg$  & \bf 3.749 (0.448)  &\bf 9.012 (1.235)   & \bf 9.742 (0.810)    &\bf 1.000 (0.000)  & 0.017(0.013)       \\
\cline{3-9} 
                           &   & \multirow{3}{*}{0.1}   & $\rm EdgeReg$ & 64.149 (1.105)  & -  &   -  & 0 &0        \\
    &   &   & $\rm DEdgeReg$ & 17.376 (0.395)  & 49.194 (7.075)  &  29.801 (1.871)   &-  & -       \\
    &   &   &  $\rm DNetReg$  & \bf 4.455 (0.390)  & \bf 10.170(1.383)   & \bf 13.701 (0.924)    &  \bf 1.000 (0.000) &    0.016 (0.011)     \\
                       \cline{2-9} 
                                               & \multirow{6}{*}{5}      & \multirow{3}{*}{0.05}  & $\rm EdgeReg$  &  59.190 (1.891)  & -  &  -   & 0 &  0      \\
                           &   &   & $\rm DEdgeReg$  & 16.418 (0.344)  & 65.339 (8.567)  & 33.260 (1.676)    &-  &  -      \\
                       
                           &   &   &  $\rm DNetReg$ &\bf  4.320 (0.503) &\bf 10.991 (1.779)   & \bf 11.374 (1.327)  & \bf 1.000 (0.000)    & 0.016 (0.011)         \\
                       \cline{3-9} 
                                               &   & \multirow{3}{*}{0.1}   &$\rm EdgeReg$  &58.110 (1.934)   & -  &  -   & 0 &  0      \\
                       
                           &   &   & $\rm DEdgeReg$  &  16.624 (0.327) & 66.029 (7.007)  & 33.824 (1.849)    &-  &-        \\
                       
                           &   &   &  $\rm DNetReg$ &\bf 5.088 (0.346)   &\bf 11.906(2.445)   & \bf 15.679 (1.423)    & \bf 1.000 (0.000) & 0.014 (0.009)        \\
                       \hline
   \end{tabular}
\end{table}

We simulate $N$ binary dynamic networks of size $n\times n$ in $[0,1]$ from model (\ref{approximatemodel}), where $\A_{jj'}(t)$, $t\in[0,1]$, follows a Bernoulli distribution and $g(\cdot)$ is taken to be the logit link function. 
The covariates $x_{i}$'s are generated independently from $\mathcal{N}(0,1)$ and we standardize the columns of the design matrix to have zero mean and unit standard deviation. 
For $\mathcal{B}_0=\sum_{r=1}^R w_{r}\u_{1r} \circ \u_{1r} \circ \u_{3r}$, we first generate the entries of $\u_{1r}$ and $\u_{3r}$ from $\mathcal{N}(0,1)$, set $w_{r}=||\u_{1r}||^2||\u_{3r}||$, and then we standardize $\u_{1r}$ and $\u_{3r}$ as unit length vectors. 
For $\mathcal{B}_1$, we randomly set $s_{0}$ proportion of its entries to be 1 and the rest to zero, such that $s_{0}=\|\mathcal{B}_1\|_0 /\left(n^{2} K\right)$. The basis functions in $\bm\phi(t)$ are set to B-spline basis with $K = 8$ equally spaced knots in $[0,1]$.

To evaluate the estimation accuracy, we report estimation errors $\|\mathcal{B}_0-\hat{\mathcal{B}}_0\|_{F}$, $\|\mathcal{B}_1-\hat{\mathcal{B}}_1\|_{F}$, and $\sum_{i=1}^{N}\|\boldsymbol{\mu}^{(i)}(t)-\hat{\boldsymbol{\mu}}^{(i)}(t)\|_{F}/N$, where $\hat{\boldsymbol{\mu}}^{(i)}(t)=g^{-1}\left(\hat{\mathcal{B}_0}\times_3 {\boldsymbol \phi}(t)+x_{i}(\hat{\mathcal{B}_1}\times_3 {\boldsymbol \phi}(t))\right)$. Furthermore, to evaluate the edge selection accuracy from our method, we report the 
true positive rate (TPR) and false positive rate (FPR) in identifying the nonzero entries in $\mathcal{B}_1$. 
The first element-wise approach $\rm EdgeReg$ does not estimate spline coefficients $\mathcal{B}_0$ and $\mathcal{B}_1$, and thus their estimation errors are not reported.
While estimates from $\rm EdgeReg$ are not sparse, the $p$-values for $\B_{ljj'}(t_h)$'s are directly available from standard GLM model fitting. In our evaluations, we apply Bonferroni correction to these p-values and then calculate the TPR and FPR in identifying the edges modulated by $x_1$, that is, entries $(j,j')$'s with nonzero time-varying covariate effects $\B_{1jj'}(t)$'s. 
Specifically, we define $\mathcal{P}^{BC} \in \mathbb{R}^{n \times n \times T}$, where $\mathcal{P}^{BC}_{jj'h}$ is the $p$-value in evaluating the significance of $\B_{1jj'}(t_h)$ from \eqref{eqn:initial0}, after the Bonferroni correction of $n \times n \times T$ tests. Defining $\bm H \in \mathbb{R}^{n \times n}$ with $\bm H_{jj'}=1{\{\min(\mathcal{P}^{BC}_{jj'.})\le 0.05\}}$, and $\bm H^{\text{true}} \in \mathbb{R}^{n \times n}$ with $\bm H^{\text{true}}_{jj'}=1{\{\int_t|\B_{1jj'}(t)|\neq 0\}}$.
The FPR and TPR are calculated as
$${\rm TPR}=\frac{\|\bm H*\bm H^{\text{true}}\|_0}{n^2s_0},\quad {\rm FPR}=\frac{\|\bm H\|_0-\|\bm H*\bm H^{\text{true}}\|_0}{n^2s_0},$$
where $*$ denotes the element-wise product. 
The second element-wise approach $\rm DEdgeReg$ does not give sparse estimates and there are no readily available inference results to calculate $p$-values, and hence their TPRs and FPRs are not reported. 

We set the number of subjects $N = 50$, the number of equally spaced time points $T=100$, and consider the number of nodes $n = 50,100$, rank $R = 2, 5$, and the sparsity proportion $s_0 = 0.05, 0.1$, respectively. 
Tables \ref{result_sim} and \ref{result_sim_100} report the average accuracy measures over 50 replications with sample size $N=50,100$, respectively, with the standard deviations shown in parentheses.
It is seen that our proposed method achieves the best performance among all competing methods, in terms of both estimation accuracy and selection accuracy, and this holds for different sample sizes $N$, numbers of nodes $n$, ranks $R$ and sparsity levels $s_0$. Moreover, the estimation error of our method $\rm DNetReg$ decreases as network size $n$, rank $R$ and sparsity proportion $s_0$ decrease, and as sample size $N$ increases. 
Estimation errors from $\rm EdgeReg$ and $\rm DEdgeReg$ are not sensitive to $R$ or $s_0$, as they are element-wise approaches and do not consider the low-rank and sparsity structure in the tensor coefficients. In terms of edge selection accuracy,  
$\rm EdgeReg$ is overly conservative after the Bonferroni correction, and its TPRs are close to zero. 
In our analysis, we also considered FDR (or BH) correction \citep{benjamini1995controlling} for $p$-value corrections and the results are similar.

\section{Application to the social cognition study in the Human Connectome Project}\label{sec:4}

The social cognition study in the HCP study collects task-related fMRI data from $N=843$ healthy adult subjects. Specifically, the fMRI data are collected on 274 evenly spaced time points covering an initiation countdown (5 seconds) followed by 5 video blocks (23 seconds each) with fixation blocks in between (15 seconds each). 
The first 11 scans in the initiation countdown period are removed in our analysis. 
The fMRI data are then preprocessed and summarized as a $68\times 263$ spatial-temporal matrix for each subject using the Desikan-Killiany Atlas \citep{desikan2006automated} with $n = 68$ ROIs (see Table \ref{68region}). For each subject, the dynamic network is constructed by calculating a sequence of connectivity matrices of dimension $68\times 68$ over $T$ sliding windows, each summarizing the connectivity between the 68 brain regions in a given window. We let the number of samples in each window and the overlap between adjacent windows be 30 and 5, respectively, giving a total of $T=47$ networks per subject. 
We determine connectivity in each individual by computing Pearson correlations between samples from a pair of regions, and create binary networks by setting $\A_{jj'}(t_h)=1$ if the computed correlation value is greater than 0.5 and $\A_{jj'}(t_h)=0$ otherwise, and this gives an average network density about 15\%. 
This procedure can eliminate weak functional connectivity and is commonly employed in existing neuroscience literature \citep{power2011functional}. 
In our analysis, we have also considered partial correlation matrices \citep{meinshausen2006high} and applied other thresholding values, such as 0.6, to the Pearson correlation matrix, and found that our main results and qualitative findings remain similar. 
In the social cognition study, there are 374 males and 469 females, aged between 22 and 36 years old. In addition, social covariates, such as companionship, social support, perceived hostility and rejection scores, are also collected for each subject. 
Our preliminary analysis finds that there are correlations between the covariates, ranging between 0.4 and 0.6.
Correspondingly, we choose to include the self-reported perceived hostility score (e.g., how often people argue with me, yell at me, or criticize me) in our analysis. 
A higher perceived hostility shows increased social distress, which is the extent to which an individual perceives his/her daily social interactions as negative or distressing \citep{lieberman2007social}. 

The goal of our analysis is to characterize the baseline brain connectivity in tasks, to ascertain how social covariates modulate the subject-level connectivity changes and to examine whether there are any sex-specific differences. We apply our proposed model to the dynamic connectivity networks from males and females, respectively. The social covariate is standardized to have mean zero and variance one, and we consider B-spline basis with $K=10$ equally spaced knots. Using the eBIC function, the rank was selected as $R=7$ and the sparsity proportion as $s_0=0.12$ for males, and $R=9$ and $s_0=0.19$ for females.



\begin{table}[!t]
	\caption{The anatomic regions of interest in the identified communities.}
	\label{community}
	\centering
	\begin{tabular}{|l|p{12cm}|}
		\hline
		1&Caudalanteriorcingulate, isthmuscingulate, paracentral, posteriorcingulate, transversetemporal, insula \\
	\hline
	2&  Cuneus, lingual, pericalcarine, postcentral, precentral, precuneus, rostralmiddlefrontal, superiorfrontal, supramarginal\\
	\hline
	3& Entorhinal, parahippocampal, temporalpole \\
	\hline
	4&Bankssts, caudalmiddlefrontal, fusiform, inferiorparietal, inferiortemporal, lateraloccipital, middletemporal, parsopercularis, parstriangularis, superiorparietal, superiortemporal\\
	\hline
	5&Lateralorbitofrontal, medialorbitofrontal, parsorbitalis, rostralanteriorcingulate, frontalpole\\
			\hline
	\end{tabular}
\end{table}

\begin{figure}[t!]
	\centering
	\subfigure[male]{
		\centering
		\includegraphics[width=0.35\linewidth]{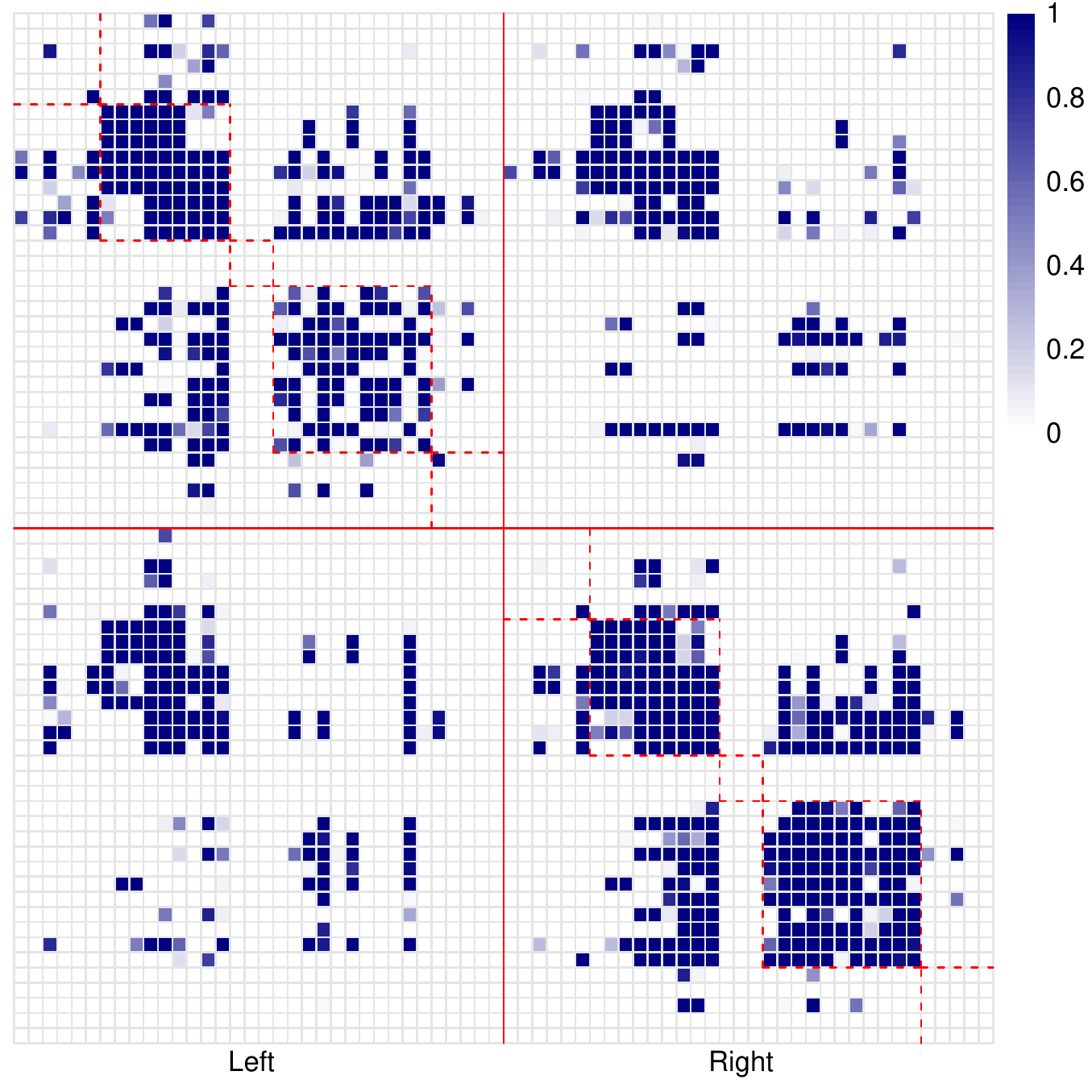}}
	\subfigure[female]{
		\centering
		\includegraphics[width=0.35\linewidth]{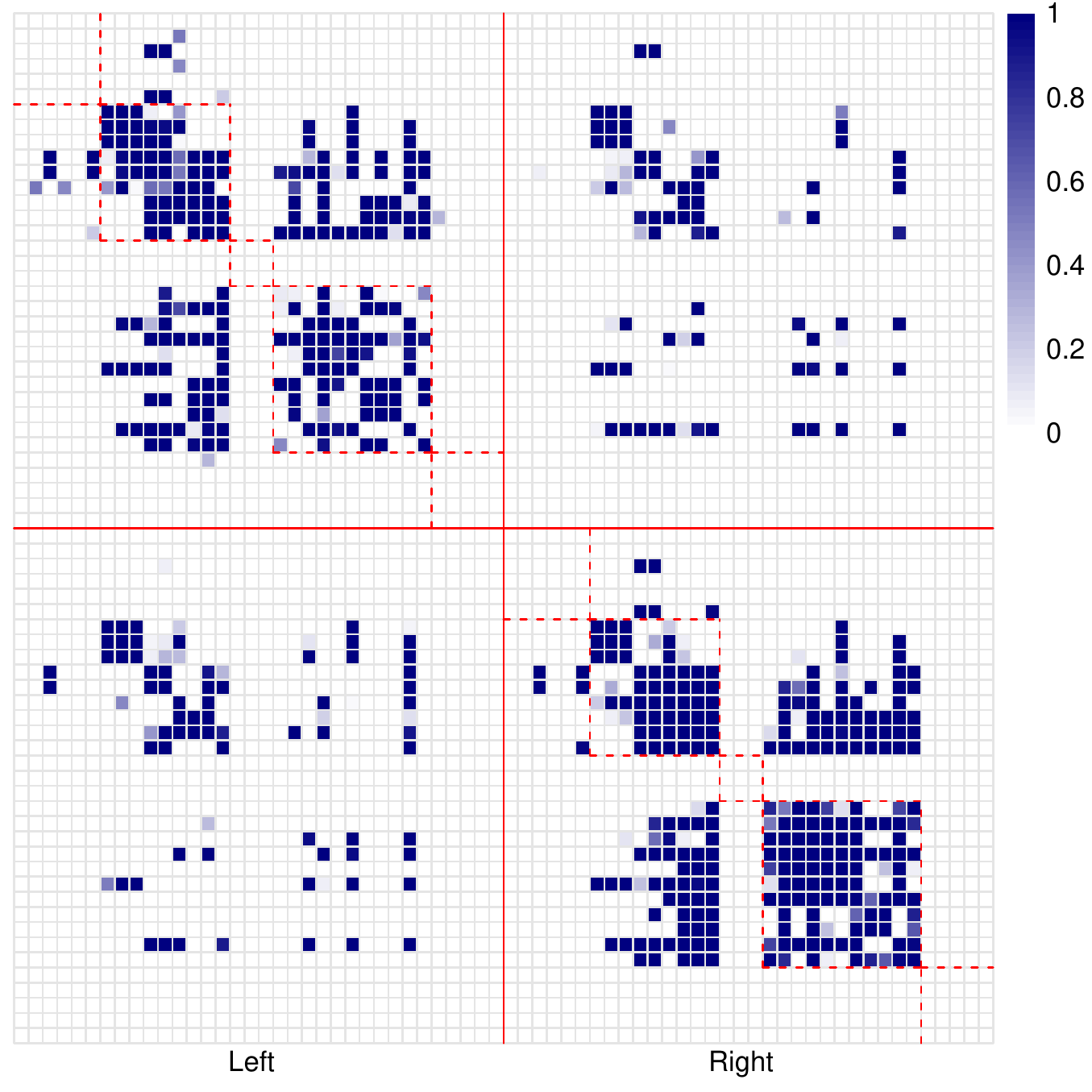}}
	\centering
	\caption{Heatmaps of the $68 \times 68$ matrix
$g^{-1}(\sum_{t=1}^T\hat{\mathcal{B}}_0\times_3 {\boldsymbol \phi}(t))$ with rows and columns ordered according to the $K$-means clustering result. Left and right hemispheres are marked in the plot. The red dashed lines mark the boundaries of the identified groups. Left and right panels are for male and female, respectively.} \label{alpha1}
\end{figure}

\smallskip\noindent
\textbf{Baseline brain connectivity.} We start by examining the estimated baseline connectivity coefficient $\hat{\mathcal B}_0$. 
Figure \ref{alpha1} plots the baseline connectivity averaged over time, i.e., $g^{-1}(\sum_{h=1}^T\mathcal{\hat B}_0\times_3 {\boldsymbol \phi}(t_h)$), where $g(\cdot)$ is the logit link function and nodes are organized by results from a K-means clustering. 
Specifically, we apply $K$-means clustering based on SVD of the average connectivity matrix $\sum_{h=1}^T\mathcal{\hat B}_0\times_3 {\boldsymbol \phi}(t_h)$ for male, and identify five clusters among the 68 ROIs. The members of each cluster are given in Table \ref{community}. 
While clustering results using $\mathcal{B}_0$ estimated for females are similar, we use the same clustering labels to facilitate comparisons.
Anatomically, the first community contains mostly nodes in the cingulate gyrus, the second and fifth communities contain nodes from the frontal lobe, the third community contains nodes from the temporal lobe, and the fourth community contains nodes from the frontal, parietal, occipital and temporal lobes (see Tables \ref{community} and \ref{68region}). 
Many of the 68 anatomic ROIs in the Desikan Atlas overlap with the resting-state functional modules. 
We find that community 1 is associated with emotion formation and processing, community 2 is related to visual, attention, and emotion regulation modules, and community 4 is enriched with visual and object identification. 
The lateral occipital gyrus in community 4, lingual gyrus in community 2, and pericalcarine gyrus in community 2 are from the occipital lobe, a region responsible for interpreting the visual world \citep{goldenberg1991contributions}, and is seen to be active for both males and females.
For both males and females, we find that connectivity between communities 2 and 4 is more active both within and between the two hemispheres, especially the temporal parietal junction, superior temporal cortex regions, and occipital gyrus, which are all relevant in social cognition. This is in line with previous research which showed that mental animations stimulate these regions \citep{castelli2000movement,barch2013function}. 
Within each hemisphere, males have higher connectivity within communities 2 and 4, 
and this is consistent with the existing findings that males have increased intrahemispheric connectivity \citep{ingalhalikar2014sex}. 

\begin{figure}[h!]
	\centering
	\subfigure[male]{
		\centering
		\includegraphics[width=0.35\linewidth]{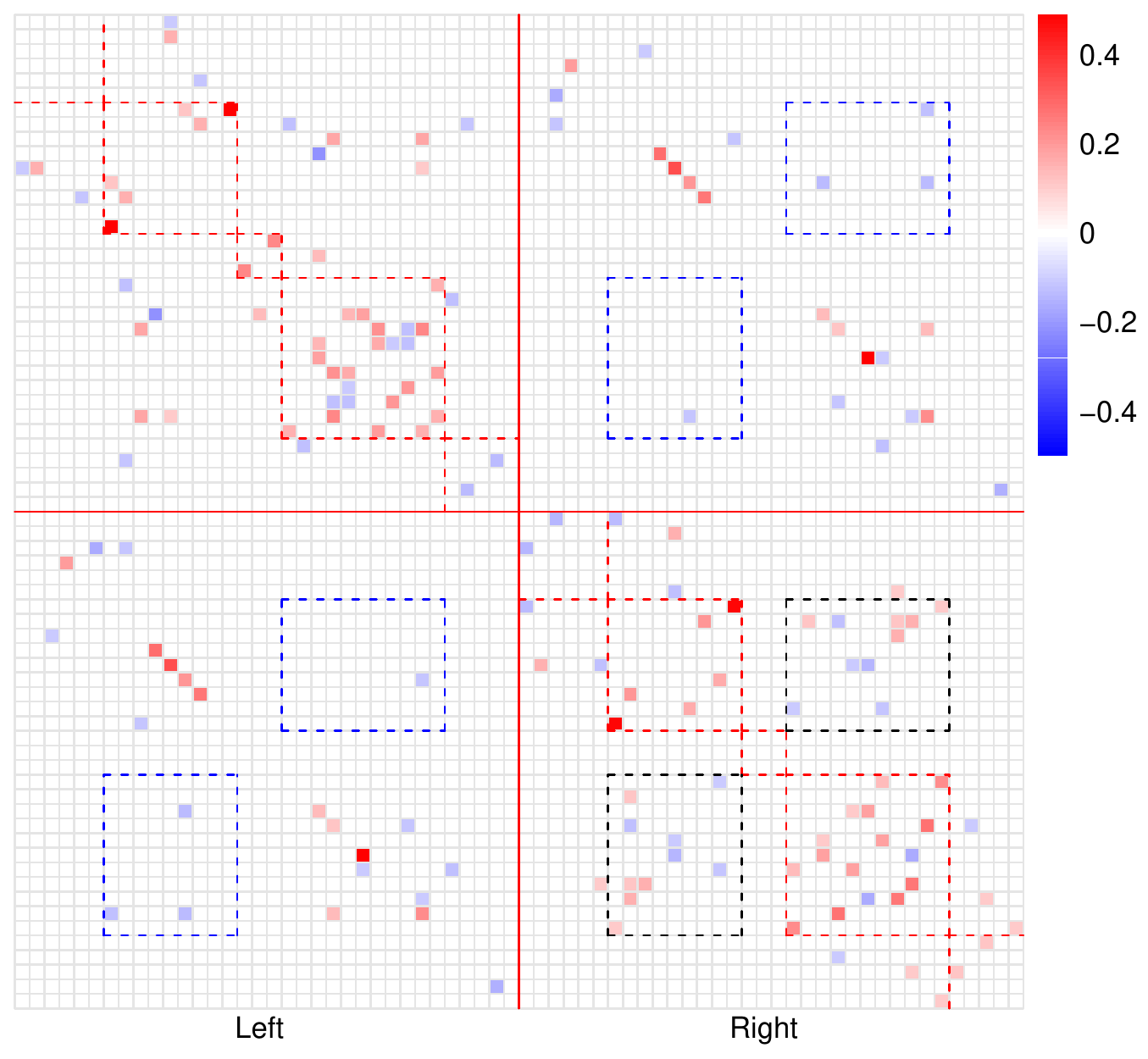}}
	\subfigure[female]{
		\centering
		\includegraphics[width=0.35\linewidth]{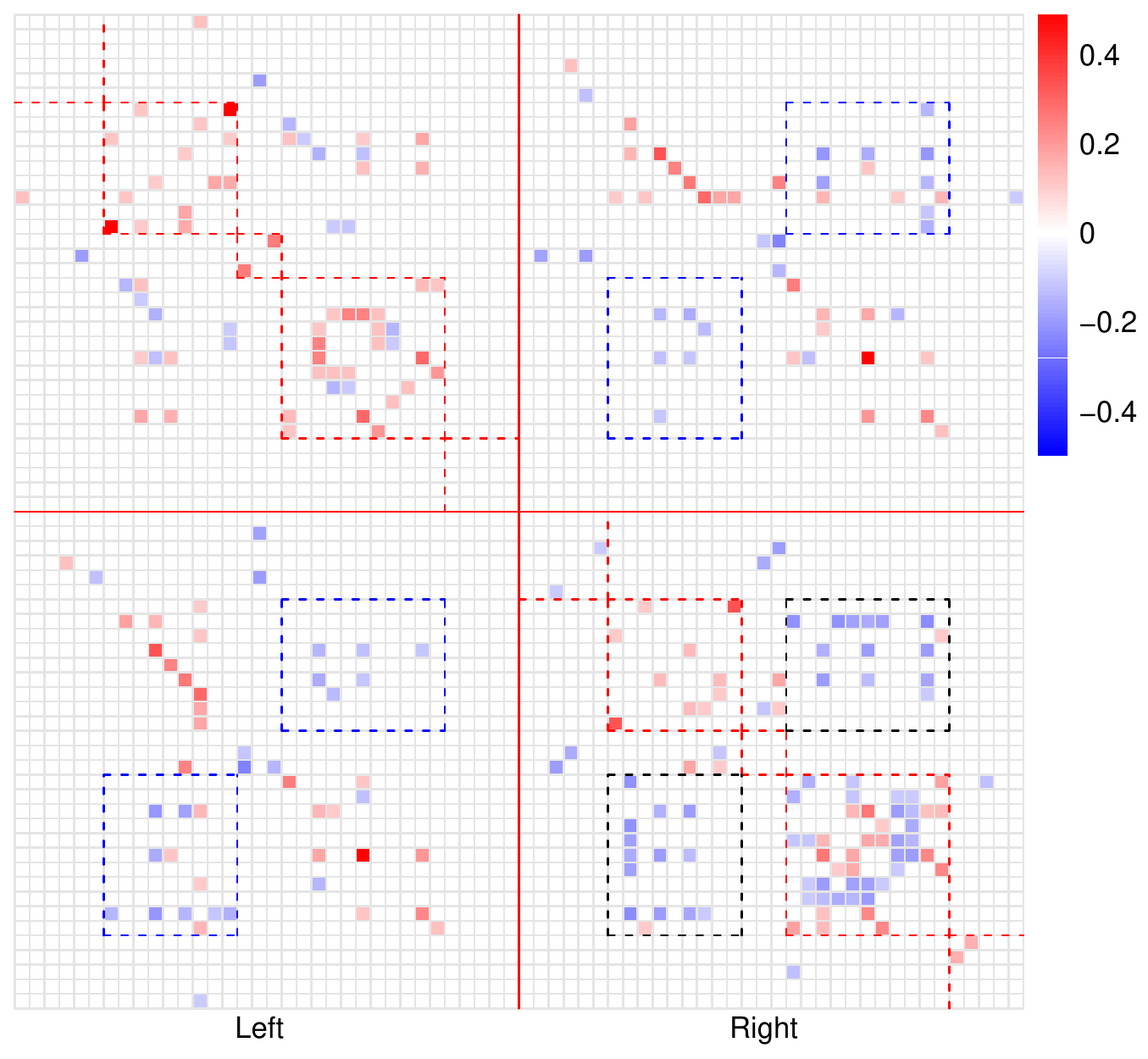}}
	\centering
	\caption{Heatmaps of $\hat{\mathcal{B}}_{1\cdot\cdot1}$ with rows and columns ordered according to the $K$-means clustering result. Left and right hemispheres are marked in the plot. The red dashed lines mark the boundaries of the identified communities within hemispheres, the black dashed lines mark the intrahemispheric connectivity between communities 2 and 4, and the blue dashed lines mark the interhemispheric connecitivity between communities 2 and 4.} \label{beta_heat}
\end{figure}

\smallskip\noindent
\textbf{Social effects on brain connectivity and sex differences.} We next examine the estimated covariate effect coefficient $\hat{\mathcal{B}}_1$. 
Figure \ref{beta_heat} plots the heatmap of estimates for males and females, where we show $\hat{\mathcal{B}}_{1\cdot\cdot1}$, the first frontal slice of $\hat{\mathcal{B}}_1$, representing the covariate effect on brain connectivity during a mental video. The values are thresholded at $\pm 0.1$ to facilitate presentation. A different view based on anatomical structure can be found in Figure \ref{beta1}.

\begin{figure}[t!]
	\centering
	\subfigure[mental video (male)]{
		\centering
		\includegraphics[width=0.3\linewidth]{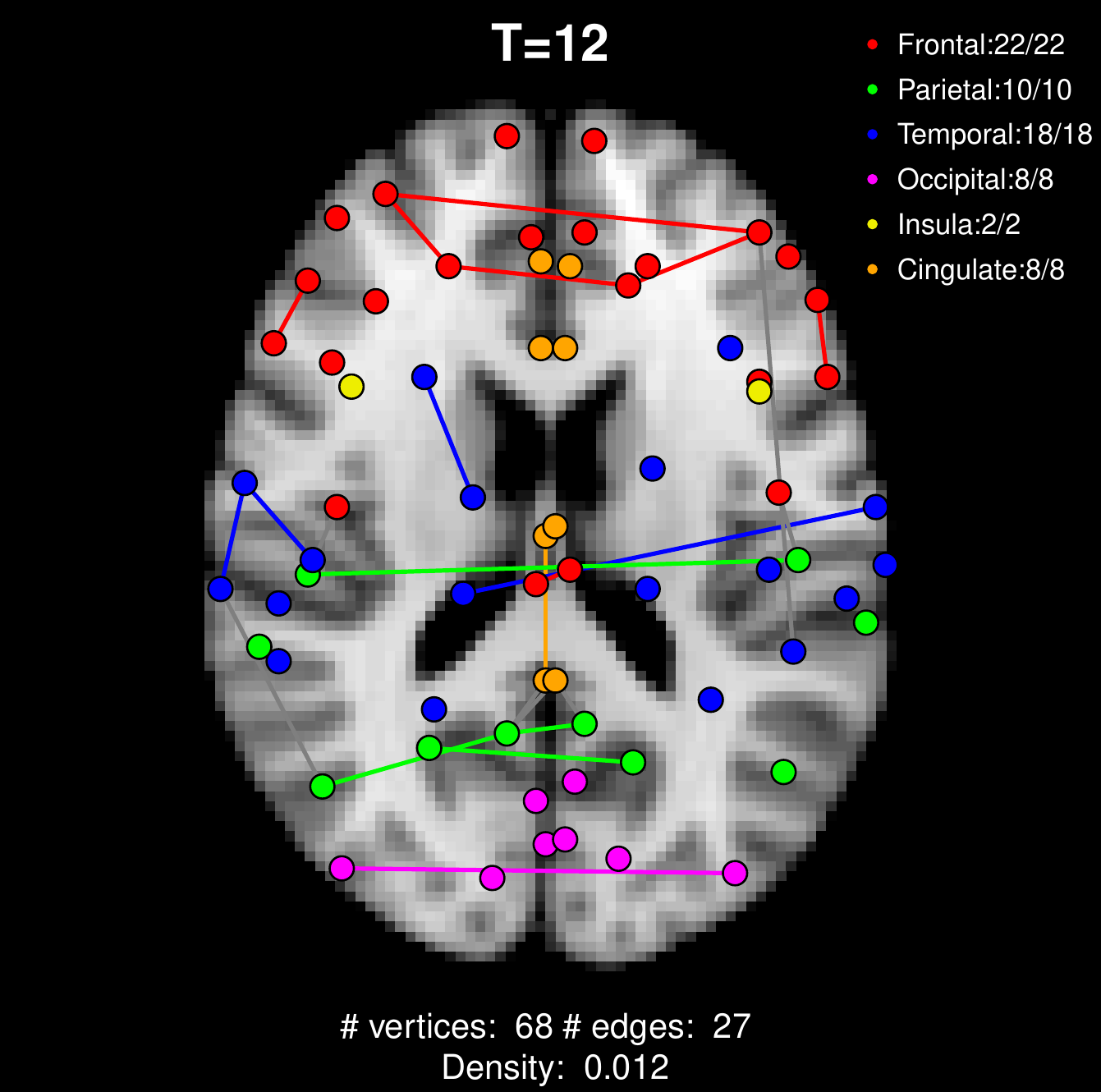}}
	\subfigure[rest (male)]{
		\centering
		\includegraphics[width=0.3\linewidth]{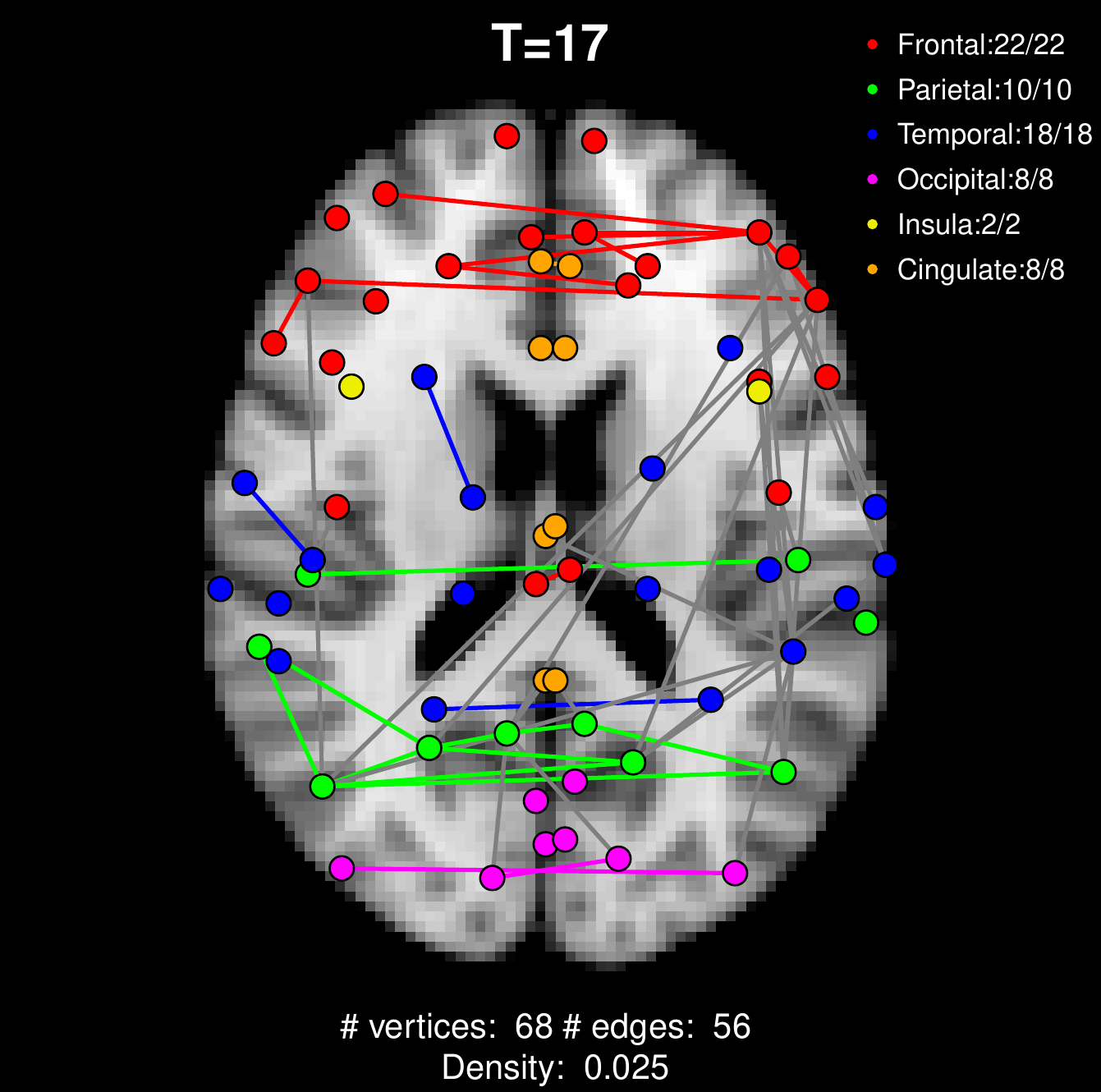}}
	\centering
		\subfigure[random video (male)]{
		\centering
		\includegraphics[width=0.3\linewidth]{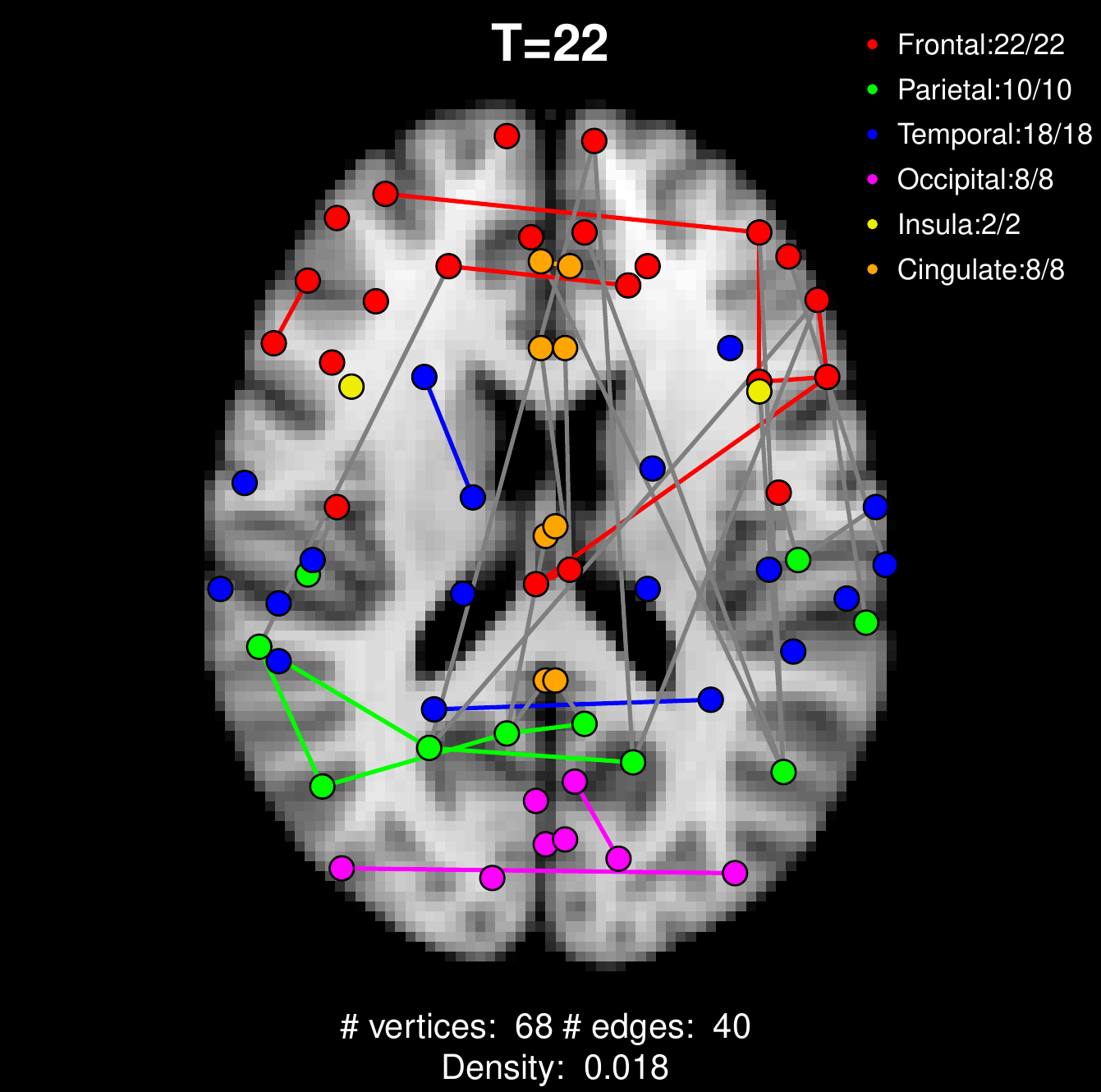}}
		\subfigure[mental video (female)]{
		\centering
		\includegraphics[width=0.3\linewidth]{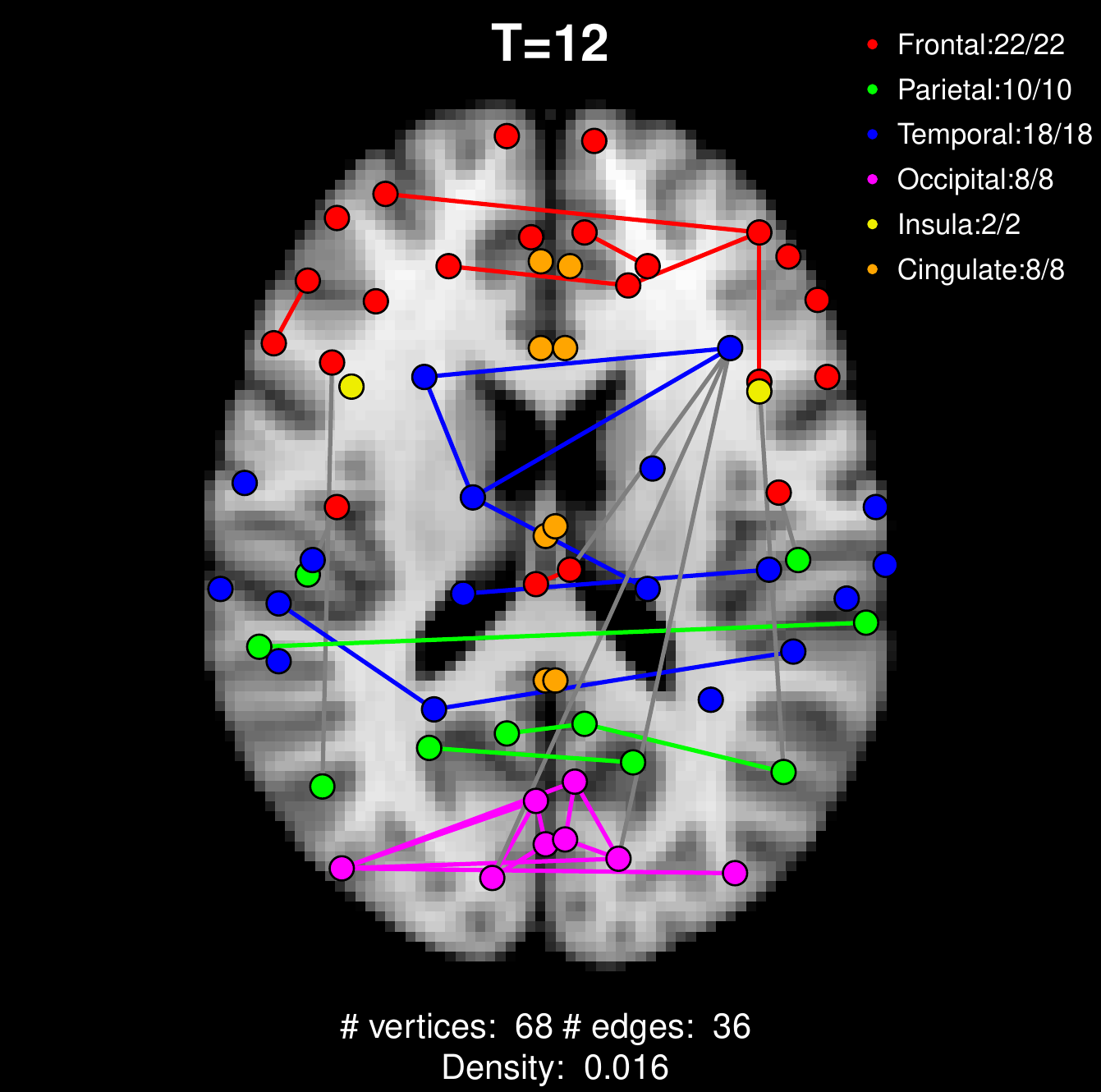}}
	\subfigure[rest (female)]{
		\centering
		\includegraphics[width=0.3\linewidth]{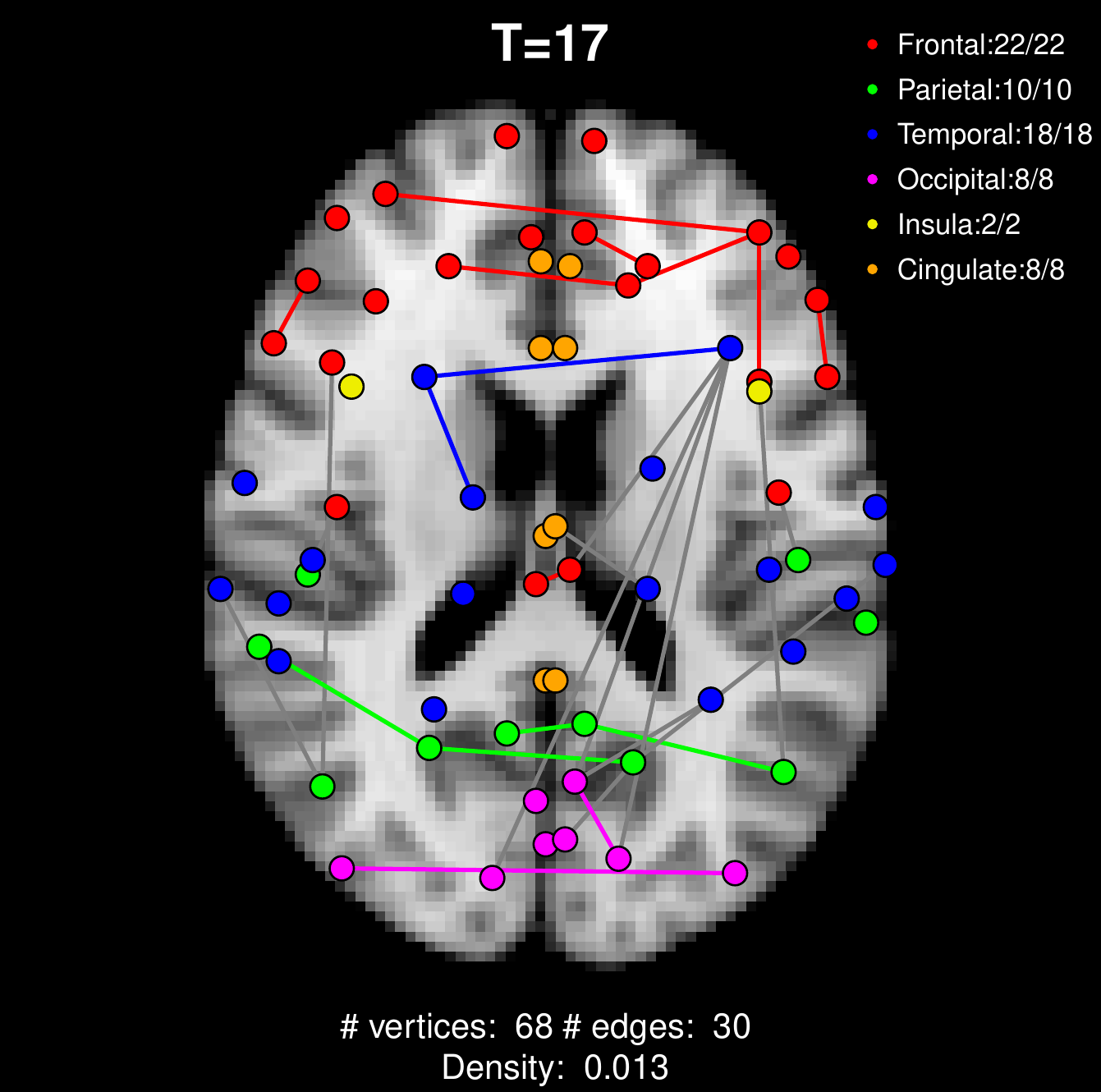}}
	\centering
		\subfigure[random video (female)]{
		\centering
		\includegraphics[width=0.3\linewidth]{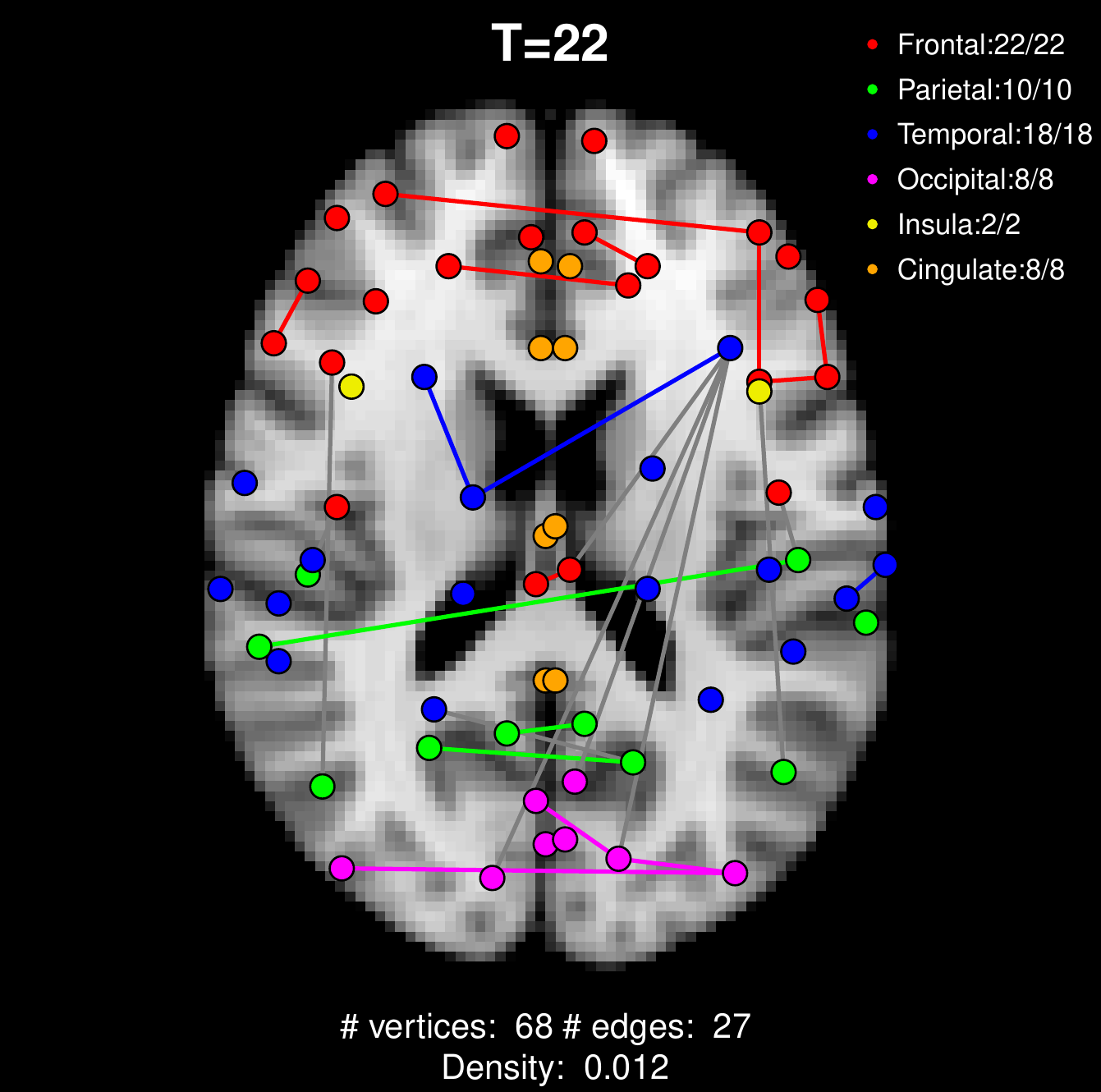}}
	\caption{Dynamic covariate effect on brain networks. The top panel plots $\hat{\mathcal{B}}_1 \times_3 {\boldsymbol \phi}(t_h)$, $h=12,17,22$, for males, and the bottom pnale plots $\hat{\mathcal{B}}_1 \times_3 {\boldsymbol \phi}(t_h)$, $h=12,17,22$, for females.} \label{beta1_trend}
\end{figure}

It is seen that the social effects on connectivity show different patterns in males and females. 
Specifically, the estimated $\hat{\mathcal{B}}_1$ has sparsity portions equal to 0.19 and 0.12 for females and males, respectively. 
Hence, the social effect on connectivity is more sparse in males, and such differences are observed in within- and between-community connectivity within and across hemispheres.
Compared to males, the social covariate is seen to more notably decrease the connectivity between communities 2 and 4 within the right hemisphere and also across hemispheres in females, suggesting that the task-related brain connectivity in females is more sensitive to social stress.
This supports existing findings that social stress influences brain connectivity and emotional perception differently for males and females \citep{mather2010sex}.
In general, the perceived hostile social distress covariate has a negative impact on the connection response for females both within and between communities, particularly for community 4, while it tends to have a positive impact on the connection response for males. The above findings on sex-specific difference are interesting, and they may be linked to existing research on sex differences in neural response to psychological stress \citep{wang2007gender}.

Finally, Figure \ref{beta1_trend} shows the social effects on brain connectivity in males and females during different periods of the experiments including watching a mental video, resting and watching a random video. 
It is seen that during a mental video, the connectivity within- and between- temporal and occipital lobes in females is more affected by social stress. The temporal lobe plays an important role in visual perception and processing emotions, and the occipital lobe is related to visual processing, containing most of the anatomical region of the visual cortex \citep{goldenberg1991contributions}. This finding suggests some interesting patterns that warrant further investigation and validation.


\subsection{A permutation based procedure to examine sex differences}\label{sec:per}
Developing the asymptotic distribution of the estimated $\mathcal{B}_1$ under the CP low-rank and sparsity constraints in our model is challenging. In this section, we conduct an ad-hoc permutation based procedure to examine whether the previously identified sex-specific differences are meaningful. 

\begin{figure}[!t]
		\centering
	\subfigure[$\D^{\text{obs}}$]{
		\centering
		\includegraphics[width=0.3\linewidth]{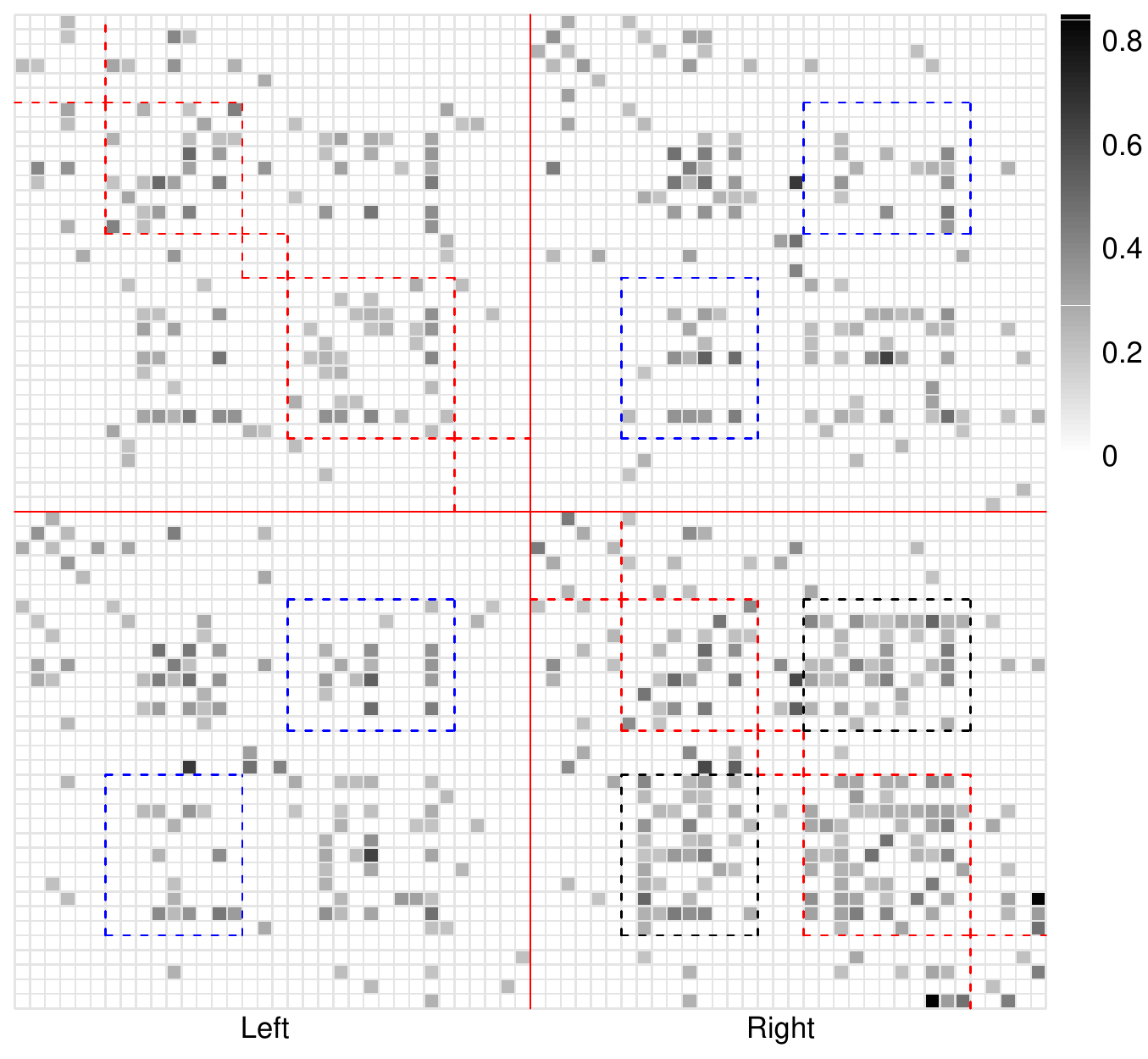}}
			\centering
	\subfigure[$\D^{\text{per}}$]{
		\centering
		\includegraphics[width=0.3\linewidth]{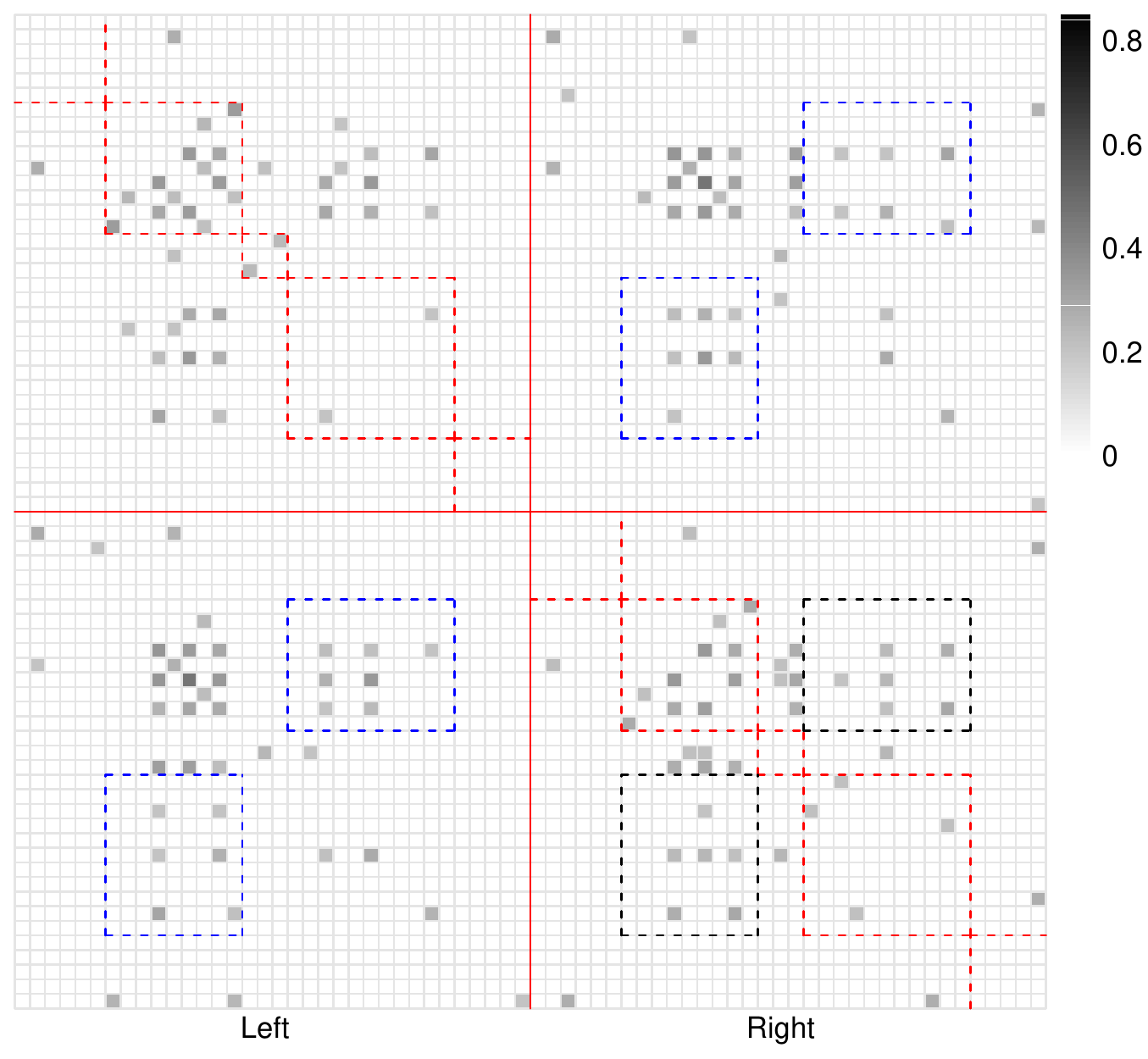}}
			\centering
			\subfigure[$\bm S$]{
		\centering
		\includegraphics[width=0.3\linewidth]{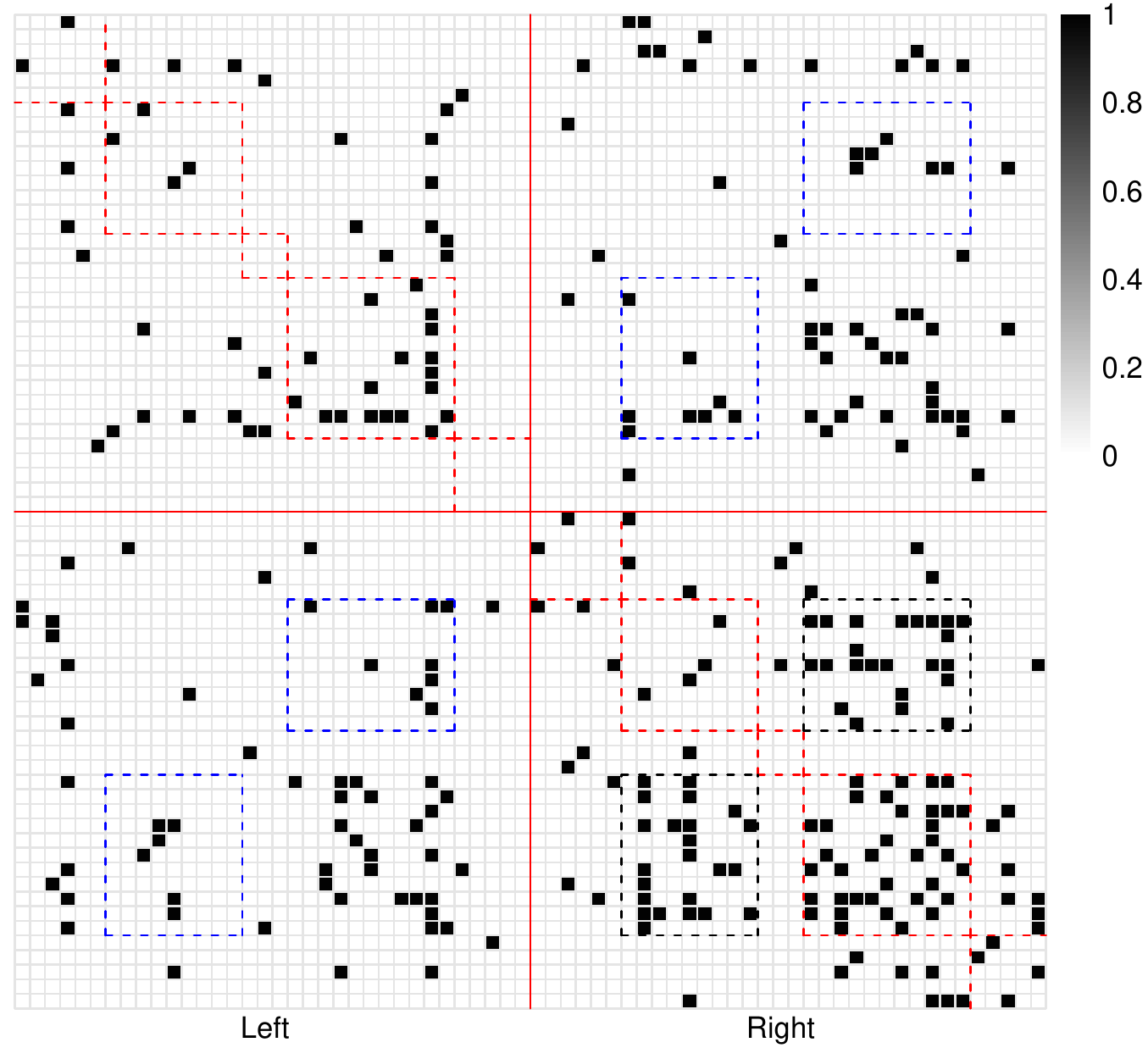}}
	\caption{Heatmaps of matrices $\D^{obs}$, $\D^{\text{per}}$ and $\bm S$.
 }\label{diff}
\end{figure}

Specifically, we randomly permute the sex labels across subjects 100 times. In each permutation $i$, we divide the $N=843$ samples into two groups based on the permuted sex labels, 
and apply the proposed model to the male and female groups, respectively.
We denote the coefficient tensors as $\mathcal{B}_0^{\text{male},i}$ (or $\mathcal{B}_0^{\text{female},i}$) and $\mathcal{B}_1^{\text{male},i}$ (or $\mathcal{B}_1^{\text{female},i}$) in permutation $i$, $i\in[100]$. 
To quantify the difference in $\mathcal{B}_1$ between males and females, we calculate the $\ell_2$ distance between the coefficient vectors for each $(j,j')$. 
Specifically, we write
\begin{equation}\label{l2dis}
\D^{obs}_{jj'}=\|\mathcal{B}^{\rm male}_{1jj'\cdot}-\mathcal{B}^{\rm female}_{1jj'\cdot}\|_2 \quad \text{and}\quad  \D^{\text{per,i}}_{jj'}=\|\mathcal{B}^{\text{ male},i}_{1jj'\cdot}-\mathcal{B}^{\text{ female},i}_{1jj'\cdot}\|_2, 
\quad j,j'\in[n],
\end{equation}
where $\mathcal{B}^{\rm male}_1, \mathcal{B}^{\rm female}_1$ are estimated based on the observed data, and 
$\mathcal{B}_1^{\text{male},i}, \mathcal{B}_1^{\text{female},i}$ are estimated based on data with the permuted sex labels. 
Figures \ref{diff} (a)-(b) show the heatmaps of $\D^{obs}$ and $\D^{\text{per}}=\sum_{i=1}^{100}\D^{\text{per,i}}/100$, respectively. 
We define a binary matrix $\bm S\in\mathbb{R}^{n\times n}$ 
$$
\bm S_{jj'}=1\left(\sum_{i=1}^{100}1(\D^{\text{obs}}_{jj'}>\D^{\text{per},i}_{jj'})\ge 95\right),
$$
where $1(\cdot)$ is the indicator function. 
Correspondingly, $\bm S_{jj'}=1$ if the observed sex difference is the same as or greater than the 95th percentile of permuted sex difference. Figure \ref{diff} (c) plots $\bm S$, which further illustrates that the sex differences within community 4 and between communities 2 and 4 are likely significant (regions in the blue and black dashed lines), affirming the findings in Figure \ref{beta_heat}. 
We also consider comparing results based on subgraphs of interests, shown in Figure \ref{boxplot}, where sex-specific differences from observed data are consistently greater than those from permuted data.

\subsection{Results using existing methods}
We evaluate the performance of two alternative methods including an elementwise method DEdgeReg, evaluated in Section \ref{sec:3}, and GLSNet \citep{zhang2022generalized}, a non time-varying matrix response regression model. Since GLSNet is not designed to model dynamic networks, we directly calculate the connectivity matrix based on all $263$ scans using the same procedure that binarizes the Pearson correlation matrix. Using GLSNet and the recommended eBIC function in \cite{zhang2022generalized}, the rank is selected as $R = 5$ and the sparsity proportion as $s_0 = 0.025$ for males, and $R=13$ and $s_0 = 0.0359$ for females. 


\begin{figure}[!t]
	\centering
	\subfigure[male (w/o thresholding)]{
		\centering
		\includegraphics[width=0.3\linewidth]{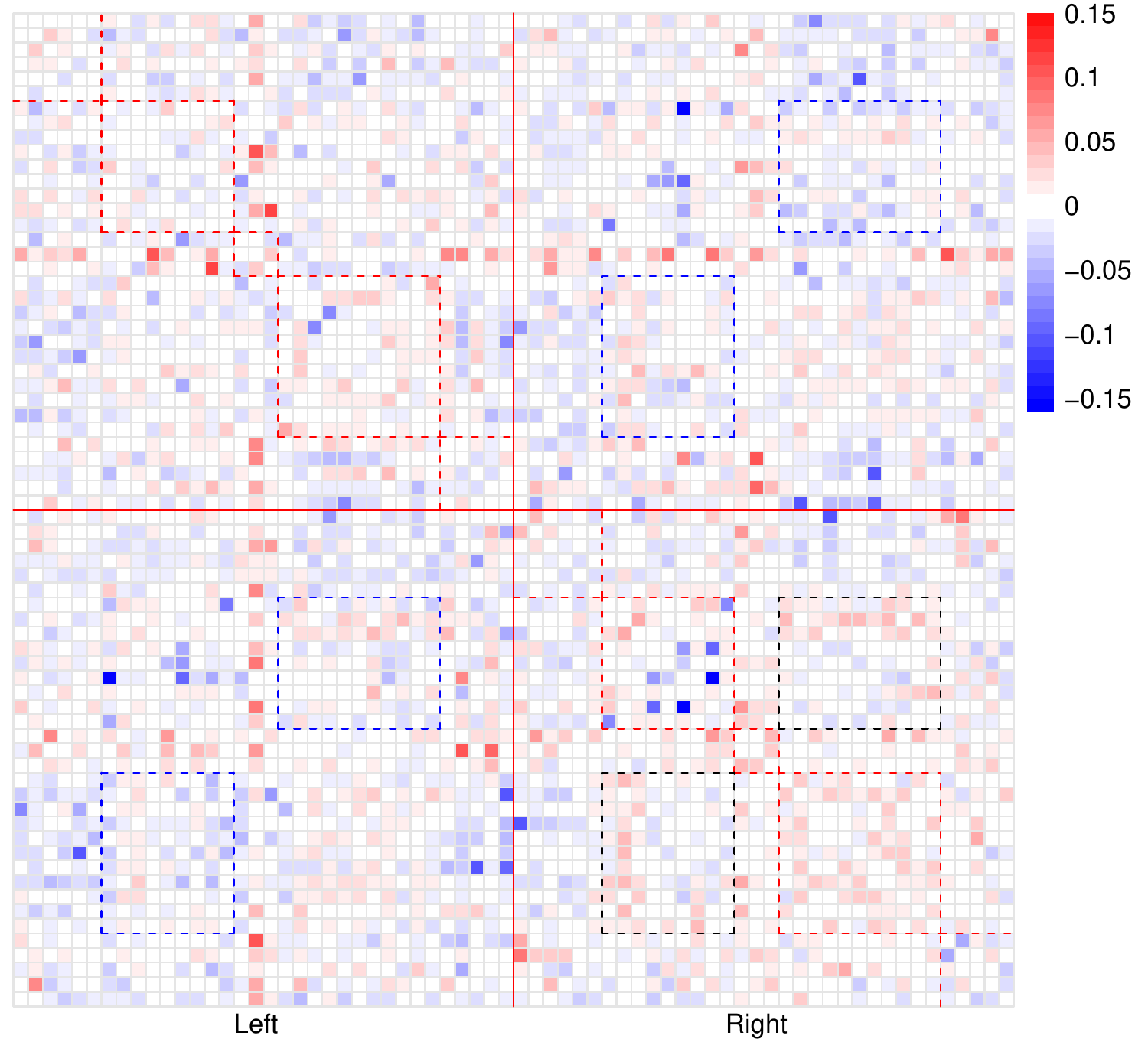}}
	\subfigure[female (w/o thresholding)]{
		\centering
		\includegraphics[width=0.3\linewidth]{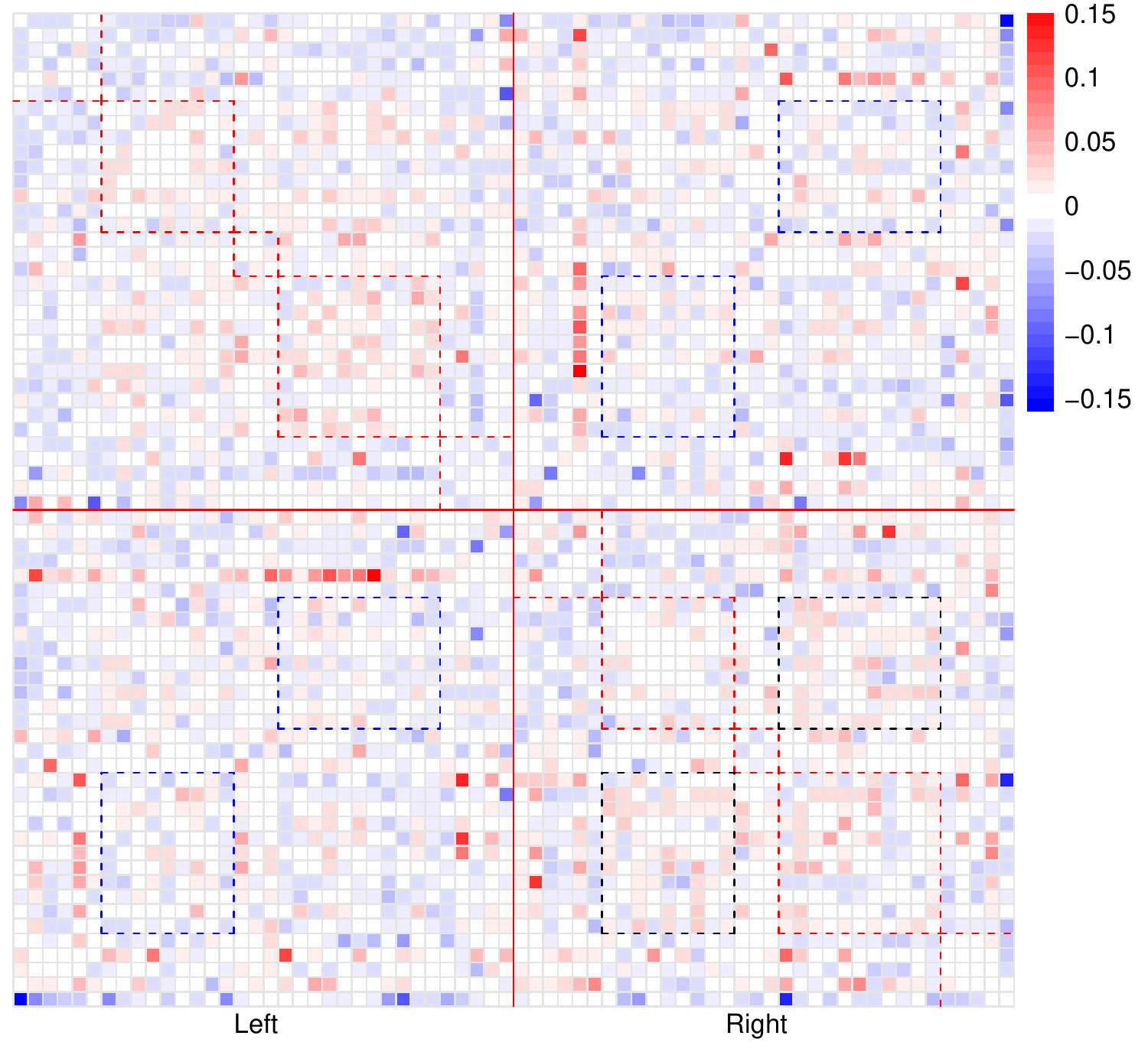}}\\
  \subfigure[male (w/ thresholding)]{
		\centering
		\includegraphics[width=0.3\linewidth]{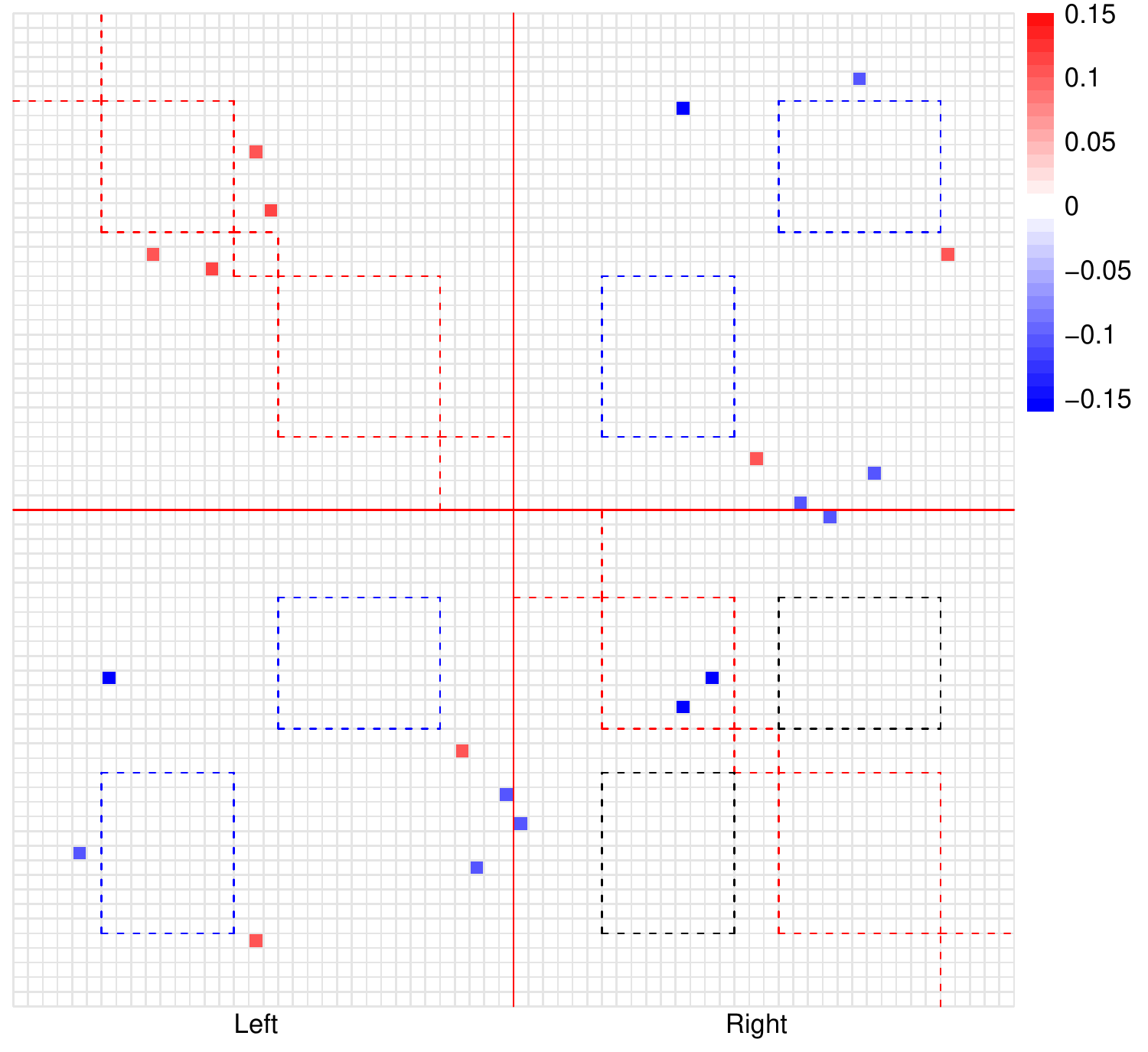}}
	\subfigure[female (w/ thresholding)]{
		\centering
		\includegraphics[width=0.3\linewidth]{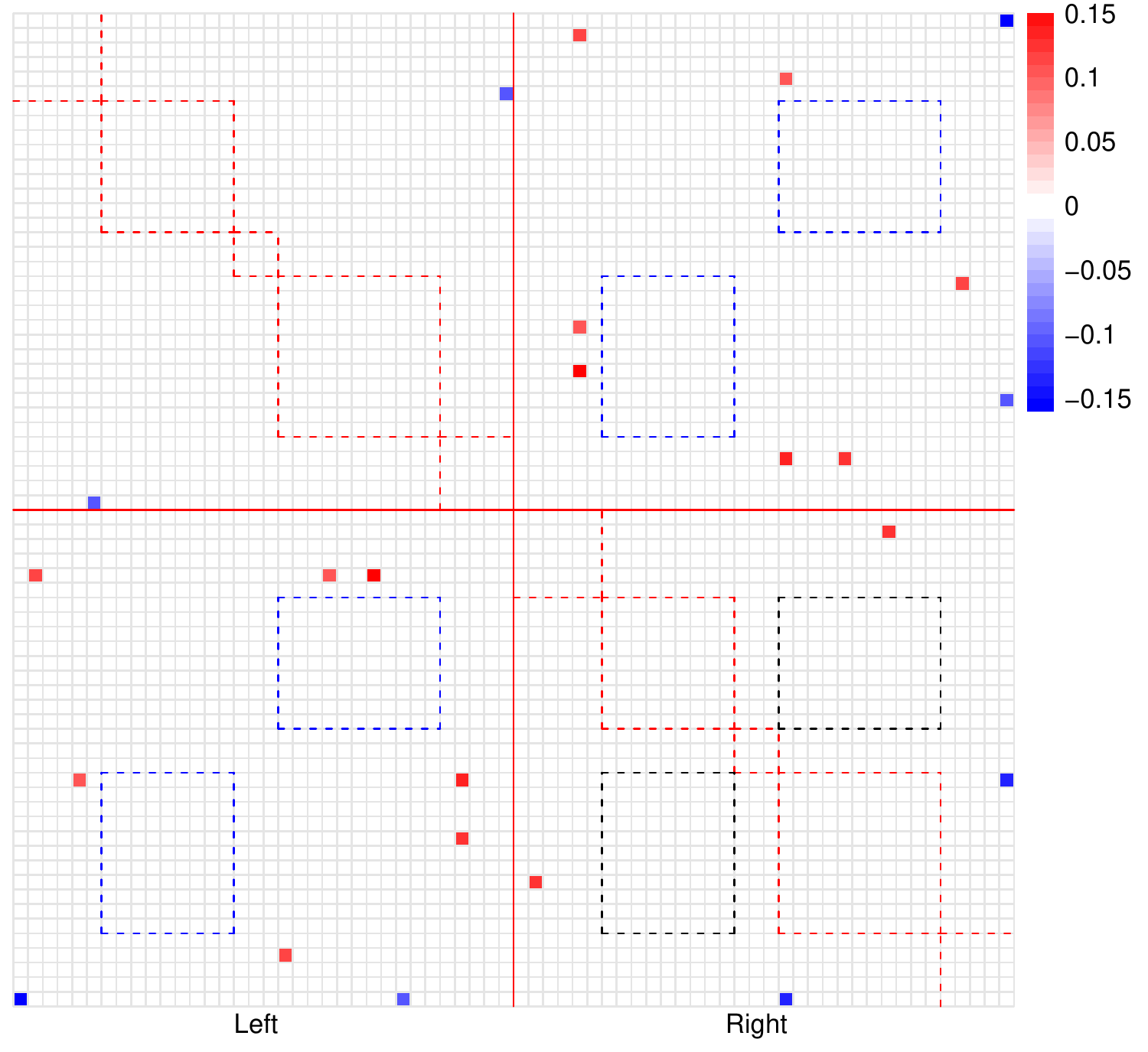}}
	\caption{Heatmaps of $\hat{\mathcal{B}}_{1\cdot\cdot1}$ estimated by DEdgeReg, with rows and columns ordered the same as Figure 4. The top and bottom panels show the results without and with thresholding, respectively.} \label{DEdgeReg}
\end{figure}

\begin{figure}[!t]
	\centering
	\subfigure[male]{
		\centering
		\includegraphics[width=0.3\linewidth]{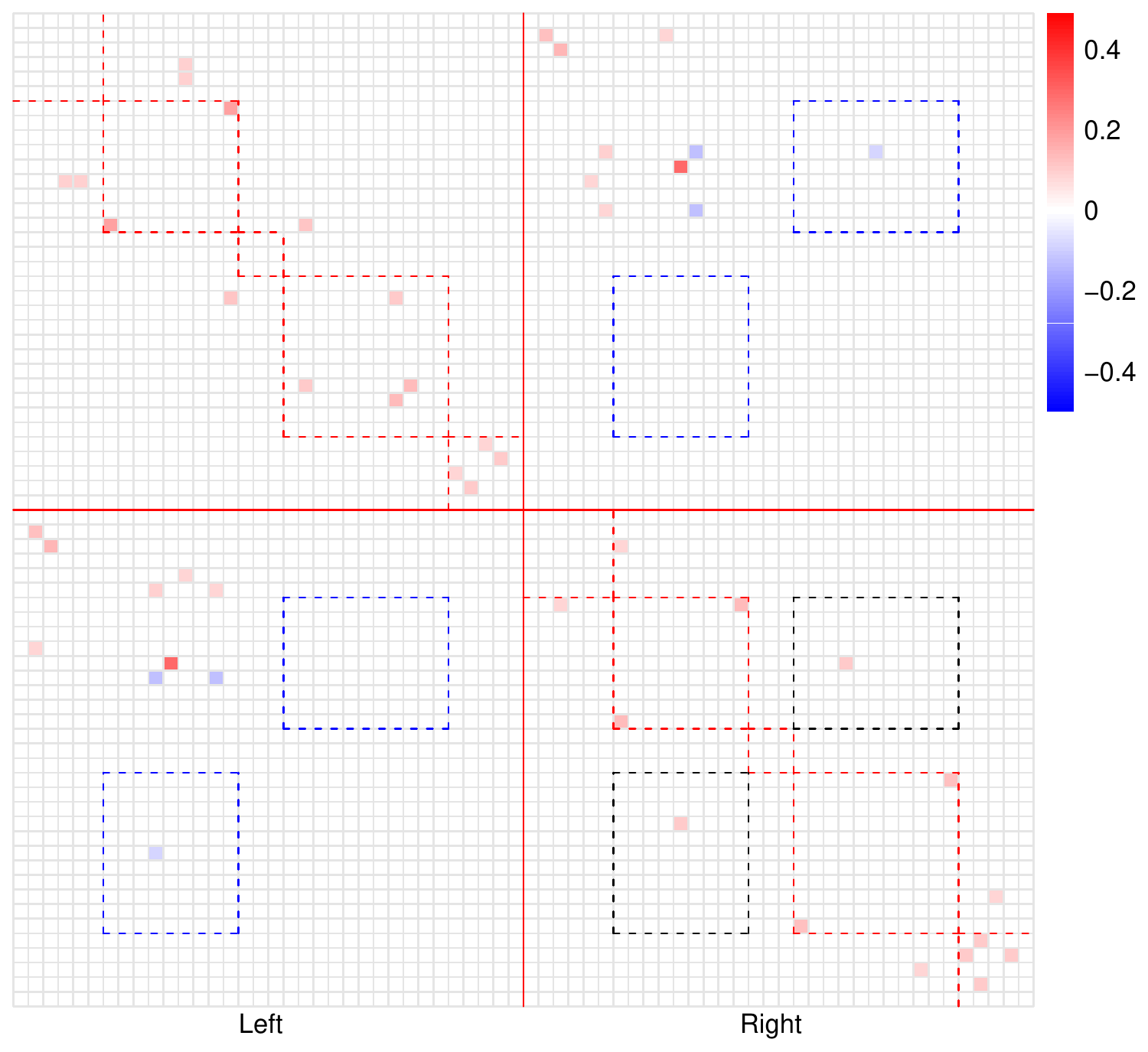}}
	\subfigure[female]{
		\centering
		\includegraphics[width=0.3\linewidth]{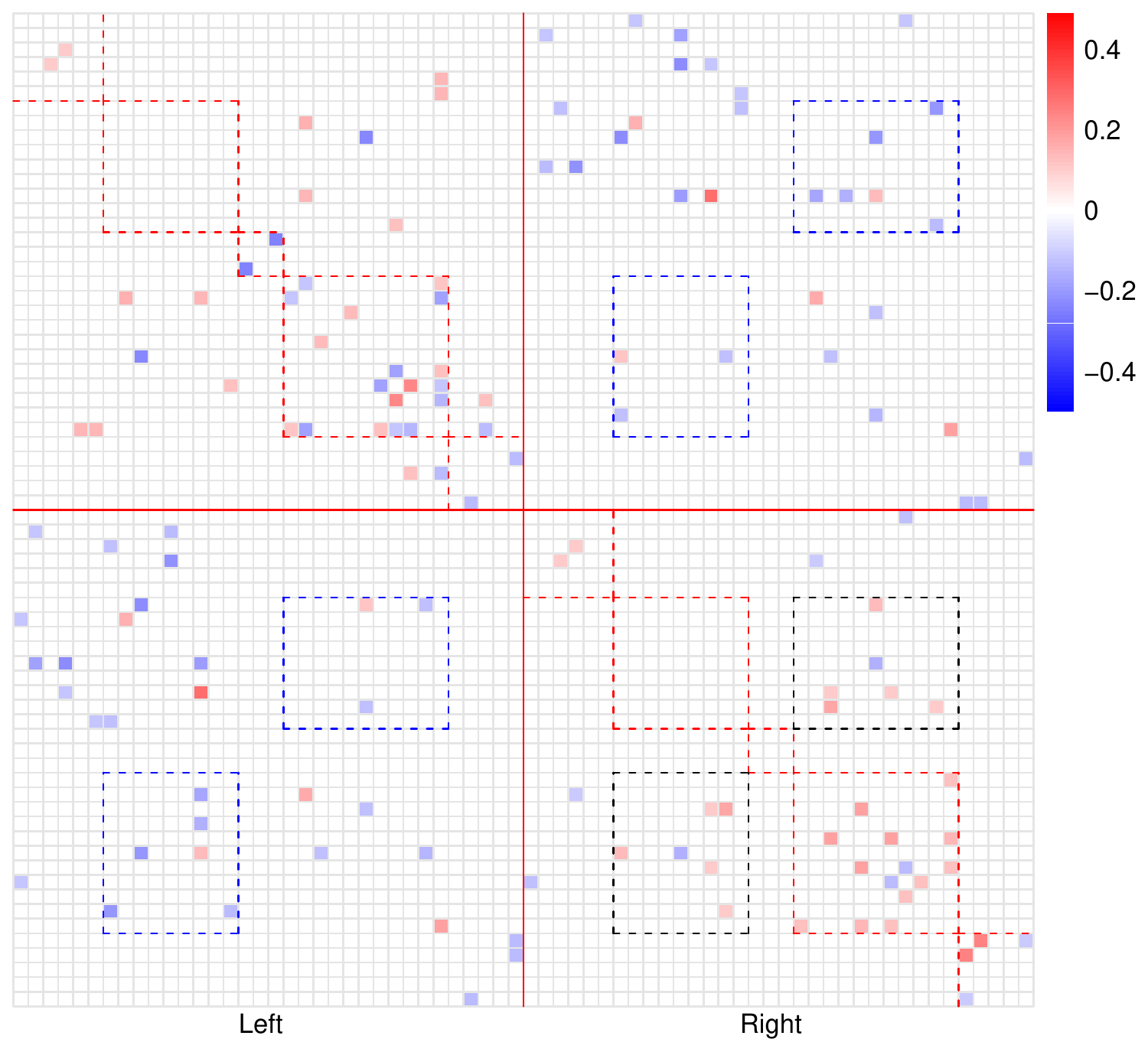}}
	\caption{Heatmaps of $\hat{{\B}}_1$ estimated by GLSNet, with rows and columns ordered the same as Figure 4. } \label{GLSNet}
\end{figure}

Figure \ref{DEdgeReg} shows $\hat{\mathcal{B}}_{1\cdot\cdot1}$ (representing the effect during a mental block) estimated by DEdgeReg with or without threshlding at $\pm 0.1$. 
It is seen that the estimates from the elementwise method DEdgeReg are very noisy and they identify a large number of regions with relatively small signals. 
The estimated social score effect coefficients $\hat{{\B}}_{1}$ from GLSNet 
are shown in Figure \ref{GLSNet}. 
For both males and females, the estimates are highly sparse.
In males, several areas associated with social cognition, such as the temporal parietal junction, superior temporal cortical regions, and occipital gyrus, do not appear to be engaged. This can potentially due to the fact that GLNet ignores the dynamic changes of brain connectivity during the experiments.

\section{Discussion}\label{sec:dis}
In this paper, we study the task-evoked brain connectivity by introducing a new semi-parametric dynamic network response regression that relates a dynamic brain connectivity network to a vector of subject-level covariates.  A key advantage of our method is to exploit the structure of dynamic
imaging coefficients in the form of high-order tensors. 
We briefly comment on potential future research. In our model setup, we assume that the tensor coefficients $\B_1,\ldots,\B_p$ are sparse. More complex structures such as the low-rank or fused structures can be considered as well, though they will increase the computation time and complexity in tuning. In Section \ref{sec:per}, we consider an ad-hoc permutation procedure to evaluate the identified sex-specific differences. A more rigorous approach would be to derive the asymptotic distribution of $\mathcal{B}_1$ and carry out hypothesis testing. This is not a trivial task due to the involvement of both low-rank and sparse constraints on the model parameters. We leave this investigation to future research.

\bibliographystyle{asa}
\bibliography{main.bbl}

\begin{thebibliography}{41}
\newcommand{\enquote}[1]{``#1''}
\expandafter\ifx\csname natexlab\endcsname\relax\def\natexlab#1{#1}\fi

\bibitem[{Adolphs(2009)}]{adolphs2009social}
Adolphs, R. (2009), \enquote{The social brain: neural basis of social
  knowledge,} \textit{Annual Review of Psychology}, 60, 693--716.

\bibitem[{Barch et~al.(2013)Barch, Burgess, Harms, Petersen, Schlaggar,
  Corbetta, Glasser, Curtiss, Dixit, Feldt, et~al.}]{barch2013function}
Barch, D.~M., Burgess, G.~C., Harms, M.~P., Petersen, S.~E., Schlaggar, B.~L.,
  Corbetta, M., Glasser, M.~F., Curtiss, S., Dixit, S., Feldt, C., et~al.
  (2013), \enquote{Function in the human connectome: task-fMRI and individual
  differences in behavior,} \textit{NeuroImage}, 80, 169--189.

\bibitem[{Beck and Teboulle(2009)}]{beck2009fast}
Beck, A. and Teboulle, M. (2009), \enquote{A fast iterative
  shrinkage-thresholding algorithm for linear inverse problems,} \textit{SIAM
  Journal on Imaging Sciences}, 2, 183--202.

\bibitem[{Benjamini and Hochberg(1995)}]{benjamini1995controlling}
Benjamini, Y. and Hochberg, Y. (1995), \enquote{Controlling the false discovery
  rate: a practical and powerful approach to multiple testing,} \textit{Journal
  of the Royal Statistical Society: Series B (Methodological)}, 57, 289--300.

\bibitem[{Bi et~al.(2018)Bi, Qu, and Shen}]{bi2018multilayer}
Bi, X., Qu, A. and Shen, X. (2018), \enquote{Multilayer tensor factorization
  with applications to recommender systems,} \textit{The Annals of Statistics},
  46, 3308--3333.

\bibitem[{Bullmore and Sporns(2009)}]{bullmore2009complex}
Bullmore, E. and Sporns, O. (2009), \enquote{Complex brain networks: graph
  theoretical analysis of structural and functional systems,} \textit{Nature
  Reviews Neuroscience}, 10, 186--198.

\bibitem[{Cai et~al.(2021)Cai, Zhang, and Sun}]{cai2021jointly}
Cai, B., Zhang, J. and Sun, W.~W. (2021), \enquote{Jointly Modeling and
  Clustering Tensors in High Dimensions,} \textit{arXiv preprint
  arXiv:2104.07773}.

\bibitem[{Castelli et~al.(2000)Castelli, Happ{\'e}, Frith, and
  Frith}]{castelli2000movement}
Castelli, F., Happ{\'e}, F., Frith, U., and Frith, C. (2000), \enquote{Movement
  and mind: a functional imaging study of perception and interpretation of
  complex intentional movement patterns,} \textit{NeuroImage}, 12, 314--325.

\bibitem[{Chen and Chen(2012)}]{chen2012extended}
Chen, J. and Chen, Z. (2012), \enquote{Extended BIC for small-n-large-P sparse
  GLM,} \textit{Statistica Sinica}, 555--574.

\bibitem[{Desikan et~al.(2006)Desikan, S{\'e}gonne, Fischl, Quinn, Dickerson,
  Blacker, Buckner, Dale, Maguire, Hyman, et~al.}]{desikan2006automated}
Desikan, R.~S., S{\'e}gonne, F., Fischl, B., Quinn, B.~T., Dickerson, B.~C.,
  Blacker, D., Buckner, R.~L., Dale, A.~M., Maguire, R.~P., Hyman, B.~T.,
  et~al. (2006), \enquote{An automated labeling system for subdividing the
  human cerebral cortex on MRI scans into gyral based regions of interest,}
  \textit{NeuroImage}, 31, 968--980.

\bibitem[{Fiske and Taylor(1991)}]{fiske1991social}
Fiske, S.~T. and Taylor, S.~E. (1991), \textit{Social cognition}, Mcgraw-Hill
  Book Company.

\bibitem[{Gallagher and Frith(2003)}]{gallagher2003functional}
Gallagher, H.~L. and Frith, C.~D. (2003), \enquote{Functional imaging of
  ‘theory of mind’,} \textit{Trends in Cognitive Sciences}, 7, 77--83.

\bibitem[{Goldenberg et~al.(1991)Goldenberg, Podreka, Steiner, Franzen, and
  Deecke}]{goldenberg1991contributions}
Goldenberg, G., Podreka, I., Steiner, M., Franzen, P., and Deecke, L. (1991),
  \enquote{Contributions of occipital and temporal brain regions to visual and
  acoustic imagery—a SPECT study,} \textit{Neuropsychologia}, 29, 695--702.

\bibitem[{Hao et~al.(2021)Hao, Wang, Wang, Zhang, Yang, and
  Sun}]{hao2021sparse}
Hao, B., Wang, B., Wang, P., Zhang, J., Yang, J., and Sun, W.~W. (2021),
  \enquote{Sparse tensor additive regression,} \textit{The Journal of Machine
  Learning Research}, 22, 2989--3031.

\bibitem[{Holland et~al.(1983)Holland, Laskey, and
  Leinhardt}]{holland1983stochastic}
Holland, P.~W., Laskey, K.~B. and Leinhardt, S. (1983), \enquote{Stochastic
  blockmodels: First steps,} \textit{Social Networks}, 5, 109--137.

\bibitem[{Hu et~al.(2021)Hu, Pan, Kong, and Shen}]{hu2021nonparametric}
Hu, W., Pan, T., Kong, D., and Shen, W. (2021), \enquote{Nonparametric matrix
  response regression with application to brain imaging data analysis,}
  \textit{Biometrics}, 77, 1227--1240.

\bibitem[{Ingalhalikar et~al.(2014)Ingalhalikar, Smith, Parker, Satterthwaite,
  Elliott, Ruparel, Hakonarson, Gur, Gur, and Verma}]{ingalhalikar2014sex}
Ingalhalikar, M., Smith, A., Parker, D., Satterthwaite, T.~D., Elliott, M.~A.,
  Ruparel, K., Hakonarson, H., Gur, R.~E., Gur, R.~C., and Verma, R. (2014),
  \enquote{Sex differences in the structural connectome of the human brain,}
  \textit{Proceedings of the National Academy of Sciences}, 111, 823--828.

\bibitem[{Kolda and Bader(2009)}]{kolda2009tensor}
Kolda, T.~G. and Bader, B.~W. (2009), \enquote{Tensor decompositions and
  applications,} \textit{SIAM Review}, 51, 455--500.

\bibitem[{Kong et~al.(2020)Kong, An, Zhang, and Zhu}]{kong2020l2rm}
Kong, D., An, B., Zhang, J., and Zhu, H. (2020), \enquote{{L2RM}: Low-rank
  linear regression models for high-dimensional matrix responses.}
  \textit{Journal of the American Statistical Association}, 115, 403--424.

\bibitem[{Lieberman(2007)}]{lieberman2007social}
Lieberman, M.~D. (2007), \enquote{Social cognitive neuroscience: a review of
  core processes,} \textit{Annual Review of Psychology}, 58, 259--289.

\bibitem[{Mather et~al.(2010)Mather, Lighthall, Nga, and
  Gorlick}]{mather2010sex}
Mather, M., Lighthall, N.~R., Nga, L., and Gorlick, M.~A. (2010), \enquote{Sex
  differences in how stress affects brain activity during face viewing,}
  \textit{Neuroreport}, 21, 933.

\bibitem[{McCullagh and Nelder(1989)}]{mccullagh1989generalized}
McCullagh, P. and Nelder, J.~A. (1989), \textit{Generalized linear models},
  vol.~37, CRC press.

\bibitem[{Meinshausen and B{\"u}hlmann(2006)}]{meinshausen2006high}
Meinshausen, N. and B{\"u}hlmann, P. (2006), \enquote{High-dimensional graphs
  and variable selection with the lasso,} .

\bibitem[{Pensky(2016)}]{pensky2016dynamic}
Pensky, M. (2016), \enquote{Dynamic network models and graphon estimation,}
  \textit{arXiv preprint arXiv:1607.00673}.

\bibitem[{Power et~al.(2011)Power, Cohen, Nelson, Wig, Barnes, Church, Vogel,
  Laumann, Miezin, Schlaggar, et~al.}]{power2011functional}
Power, J.~D., Cohen, A.~L., Nelson, S.~M., Wig, G.~S., Barnes, K.~A., Church,
  J.~A., Vogel, A.~C., Laumann, T.~O., Miezin, F.~M., Schlaggar, B.~L., et~al.
  (2011), \enquote{Functional network organization of the human brain,}
  \textit{Neuron}, 72, 665--678.

\bibitem[{Saxe and Kanwisher(2013)}]{saxe2013people}
Saxe, R. and Kanwisher, N. (2013), \enquote{People thinking about thinking
  people: the role of the temporo-parietal junction in “theory of mind”,}
  in \textit{Social Neuroscience}, Psychology Press, pp. 171--182.

\bibitem[{Schnabel et~al.(1985)Schnabel, Koonatz, and
  Weiss}]{schnabel1985modular}
Schnabel, R.~B., Koonatz, J.~E. and Weiss, B.~E. (1985), \enquote{A modular
  system of algorithms for unconstrained minimization,} \textit{ACM
  Transactions on Mathematical Software (TOMS)}, 11, 419--440.

\bibitem[{Sch{\"o}lvinck et~al.(2010)Sch{\"o}lvinck, Maier, Ye, Duyn, and
  Leopold}]{scholvinck2010neural}
Sch{\"o}lvinck, M.~L., Maier, A., Ye, F.~Q., Duyn, J.~H., and Leopold, D.~A.
  (2010), \enquote{Neural basis of global resting-state fMRI activity,}
  \textit{Proceedings of the National Academy of Sciences}, 107, 10238--10243.

\bibitem[{Smith et~al.(2013)Smith, Vidaurre, Beckmann, Glasser, Jenkinson,
  Miller, Nichols, Robinson, Salimi-Khorshidi, Woolrich,
  et~al.}]{smith2013functional}
Smith, S.~M., Vidaurre, D., Beckmann, C.~F., Glasser, M.~F., Jenkinson, M.,
  Miller, K.~L., Nichols, T.~E., Robinson, E.~C., Salimi-Khorshidi, G.,
  Woolrich, M.~W., et~al. (2013), \enquote{Functional connectomics from
  resting-state fMRI,} \textit{Trends in Cognitive Sciences}, 17, 666--682.

\bibitem[{Srivastava et~al.(2017)Srivastava, Engelhardt, and
  Dunson}]{srivastava2017expandable}
Srivastava, S., Engelhardt, B.~E. and Dunson, D.~B. (2017), \enquote{Expandable
  factor analysis,} \textit{Biometrika}, 104, 649--663.

\bibitem[{Tang et~al.(2020)Tang, Bi, and Qu}]{tang2020individualized}
Tang, X., Bi, X. and Qu, A. (2020), \enquote{Individualized multilayer tensor
  learning with an application in imaging analysis,} \textit{Journal of the
  American Statistical Association}, 115, 836--851.

\bibitem[{Wang et~al.(2007)Wang, Korczykowski, Rao, Fan, Pluta, Gur, McEwen,
  and Detre}]{wang2007gender}
Wang, J., Korczykowski, M., Rao, H., Fan, Y., Pluta, J., Gur, R.~C., McEwen,
  B.~S., and Detre, J.~A. (2007), \enquote{Gender difference in neural response
  to psychological stress,} \textit{Social cognitive and affective
  neuroscience}, 2, 227--239.

\bibitem[{Wang et~al.(2017)Wang, Durante, Jung, and Dunson}]{wang2017bayesian}
Wang, L., Durante, D., Jung, R.~E., and Dunson, D.~B. (2017), \enquote{Bayesian
  network-response regression,} \textit{Bioinformatics}, 33, 1859--1866.

\bibitem[{Wheatley et~al.(2007)Wheatley, Milleville, and
  Martin}]{wheatley2007understanding}
Wheatley, T., Milleville, S.~C. and Martin, A. (2007), \enquote{Understanding
  animate agents: distinct roles for the social network and mirror system,}
  \textit{Psychological Science}, 18, 469--474.

\bibitem[{Xu and Hero(2014)}]{xu2014dynamic}
Xu, K.~S. and Hero, A.~O. (2014), \enquote{Dynamic stochastic blockmodels for
  time-evolving social networks,} \textit{IEEE Journal of Selected Topics in
  Signal Processing}, 8, 552--562.

\bibitem[{Yuan and Lin(2006)}]{yuan2006model}
Yuan, M. and Lin, Y. (2006), \enquote{Model selection and estimation in
  regression with grouped variables,} \textit{Journal of the Royal Statistical
  Society: Series B (Statistical Methodology)}, 68, 49--67.

\bibitem[{Zhang and Cao(2017)}]{ZhangCao2017}
Zhang, J. and Cao, J. (2017), \enquote{Finding Common Modules in a Time-Varying
  Network with Application to the Drosophila Melanogaster Gene Regulation
  Network,} \textit{Journal of the American Statistical Association}, 112,
  994--1008.

\bibitem[{Zhang et~al.(2020)Zhang, Sun, and Li}]{zhang2020mixed}
Zhang, J., Sun, W.~W. and Li, L. (2020), \enquote{Mixed-effect time-varying
  network model and application in brain connectivity analysis,}
  \textit{Journal of the American Statistical Association}, 115, 2022--2036.

\bibitem[{Zhang et~al.(2023)Zhang, Sun, and Li}]{zhang2022generalized}
Zhang, J., Sun, W.~W. and Li, L. (2023),, \enquote{Generalized Connectivity Matrix Response Regression with
  Applications in Brain Connectivity Studies,} \textit{Journal of Computational
  and Graphical Statistics}, 32, 252--262.

\bibitem[{Zhang and Li(2017)}]{zhang2017tensor}
Zhang, X. and Li, L. (2017), \enquote{Tensor envelope partial least-squares
  regression,} \textit{Technometrics}, 59, 426--436.

\bibitem[{Zhou et~al.(2021)Zhou, Sun, Zhang, and Li}]{zhou2021partially}
Zhou, J., Sun, W.~W., Zhang, J., and Li, L. (2021), \enquote{Partially observed
  dynamic tensor response regression,} \textit{Journal of the American
  Statistical Association}, 1--16.

\end{thebibliography}

\newpage
\renewcommand{\thesection}{S\arabic{section}}
\renewcommand{\theequation}{S\arabic{equation}}
\renewcommand{\thefigure}{S\arabic{figure}}
\renewcommand{\thetable}{S\arabic{table}}
\setcounter{section}{0}
\setcounter{equation}{0}
\setcounter{figure}{0}
\setcounter{table}{0}
\setcounter{page}{1}

\begin{center}
{\large\bf Supplementary Materials for ``Learning Brain Connectivity in Social Cognition with Dynamic Network Regression"} \\
\bigskip
\end{center}

\section{Gradients}\label{sec:gr}
We present the analytical forms of the gradients in Algorithm 1:
$$
\begin{aligned}
\ell(\mathcal{B}_0,\ldots,\mathcal{B}_p)&=-\frac{1}{N}\sum^N_{i=1}\sum^n_{j\neq j'}\sum^T_{h=1}\Big[\A^{(i)}_{jj'}(t_h)\bfeta^{(i)}_{jj'}(t_h)-\psi\left\{\bfeta^{(i)}_{jj'}(t_h)\right\}\Big]\\
&=-\frac{1}{N}\sum^N_{i=1}\sum^n_{j\neq j'}\sum^T_{h=1}\Big[\A^{(i)}(t_h)*\left\{\mathcal{B}_0\times_3 {\boldsymbol \phi}(t)+\sum_{l=1}^p x_{il}(\mathcal{B}_l\times_3 {\boldsymbol \phi}(t))\right\}\\
&-\log\left(1+\exp\left\{\mathcal{B}_0\times_3 {\boldsymbol \phi}(t)+\sum_{l=1}^p x_{il}(\mathcal{B}_l\times_3 {\boldsymbol \phi}(t))\right\}\right)\Big],\\
\end{aligned}
$$

$$
\frac{\partial l}{\partial u_{1r}}=-\frac{1}{N}\sum_{i=1}^N\sum_{t=1}^T\left(A^{(i)}(t)-{\boldsymbol \psi}'\left(\mathcal{B}_0\times_3{\boldsymbol \phi}(t)+\sum_{l=1}^px_{il}\mathcal{B}_l\times_3{\boldsymbol \phi}(t)\right)\right)\left(w_ru_{1r}{\boldsymbol \phi}^T(t)u_{3r}\right),
$$

$$
\frac{\partial l}{\partial u_{3r}}=-\frac{1}{N}\sum_{i=1}^N\sum_{t=1}^T\left(A^{(i)}(t)-{\boldsymbol \psi}'\left(\mathcal{B}_0\times_3{\boldsymbol \phi}(t)+\sum_{l=1}^px_{il}\mathcal{B}_l\times_3{\boldsymbol \phi}(t)\right)\right)\left(w_ru_{1r}{\boldsymbol \phi}^T(t)u_{1r}\right),
$$

$$
\frac{\partial l}{\partial\mathcal{B}_l }=-\frac{1}{N}\sum_{i=1}^N\sum_{t=1}^T\left\{A^{(i)}(t)\circ\left(x_{il}{\boldsymbol \phi}(t)\right)-{\boldsymbol \psi}'\left(\mathcal{B}_0\times_3{\boldsymbol \phi}(t)+\sum_{l=1}^px_{il}\mathcal{B}_l\times_3{\boldsymbol \phi}(t)\right)\circ\left(x_{il}{\boldsymbol \phi}(t)\right)\right\},
$$
where ${\boldsymbol \psi}'(x)=1-1/(1+\exp(x))$.

\section{Additional results from real data analysis}

\begin{figure}[t!]
			\centering
		\subfigure[whole]{
		\centering
		\includegraphics[width=0.3\linewidth]{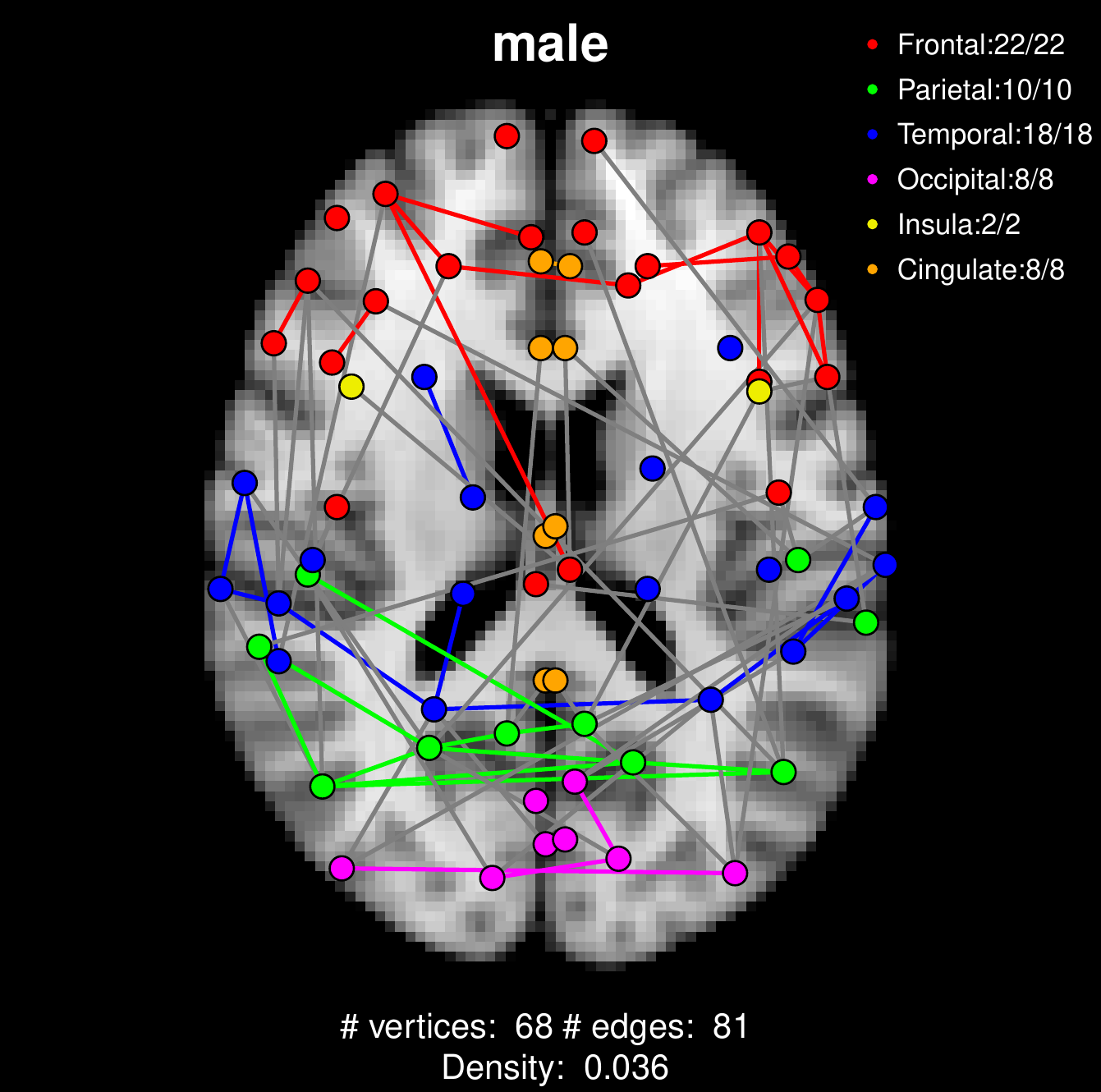}}
			\centering
		\subfigure[left]{
		\centering
		\includegraphics[width=0.31\linewidth]{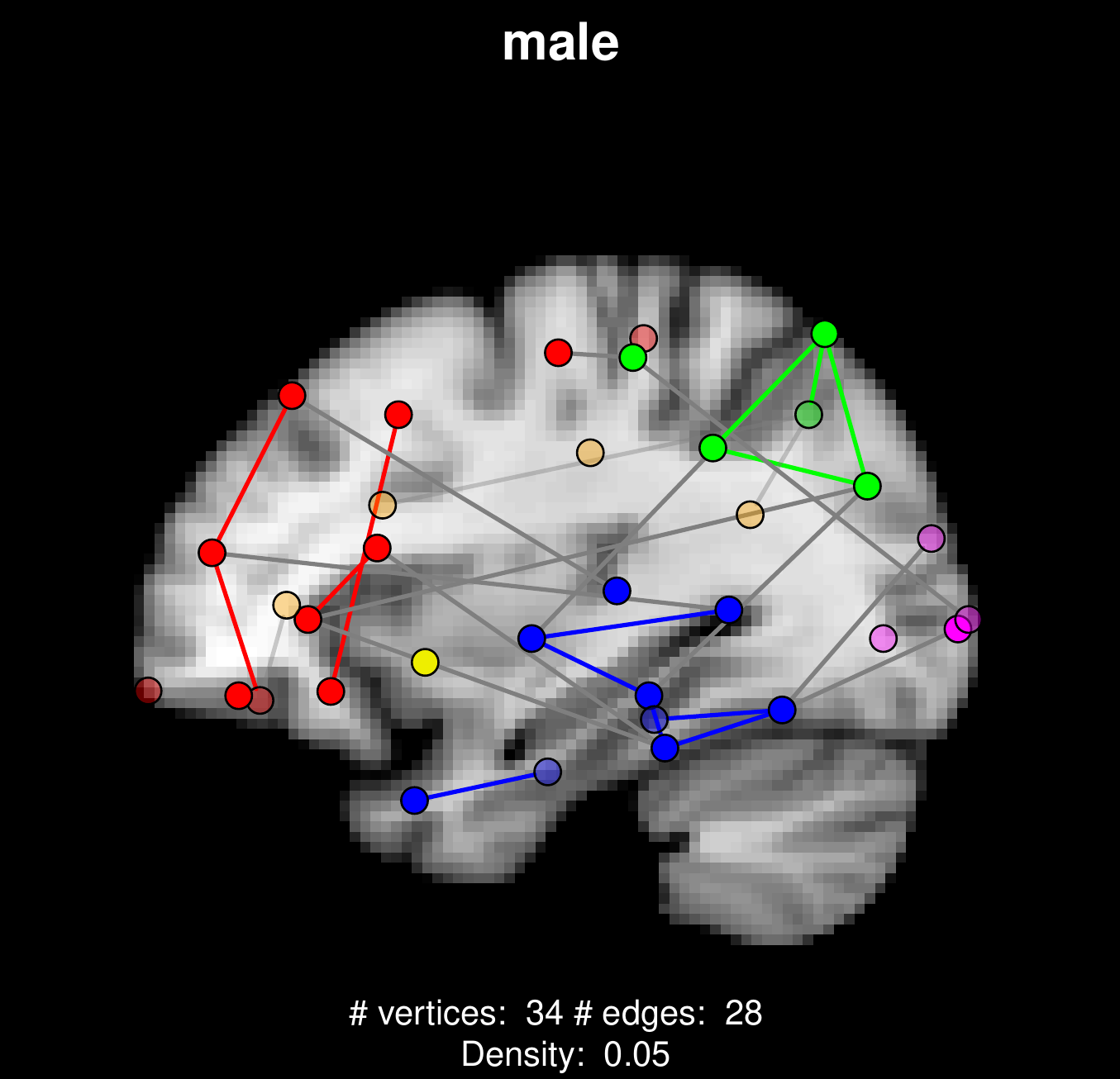}}
			\centering
		\subfigure[right]{
		\centering
		\includegraphics[width=0.31\linewidth]{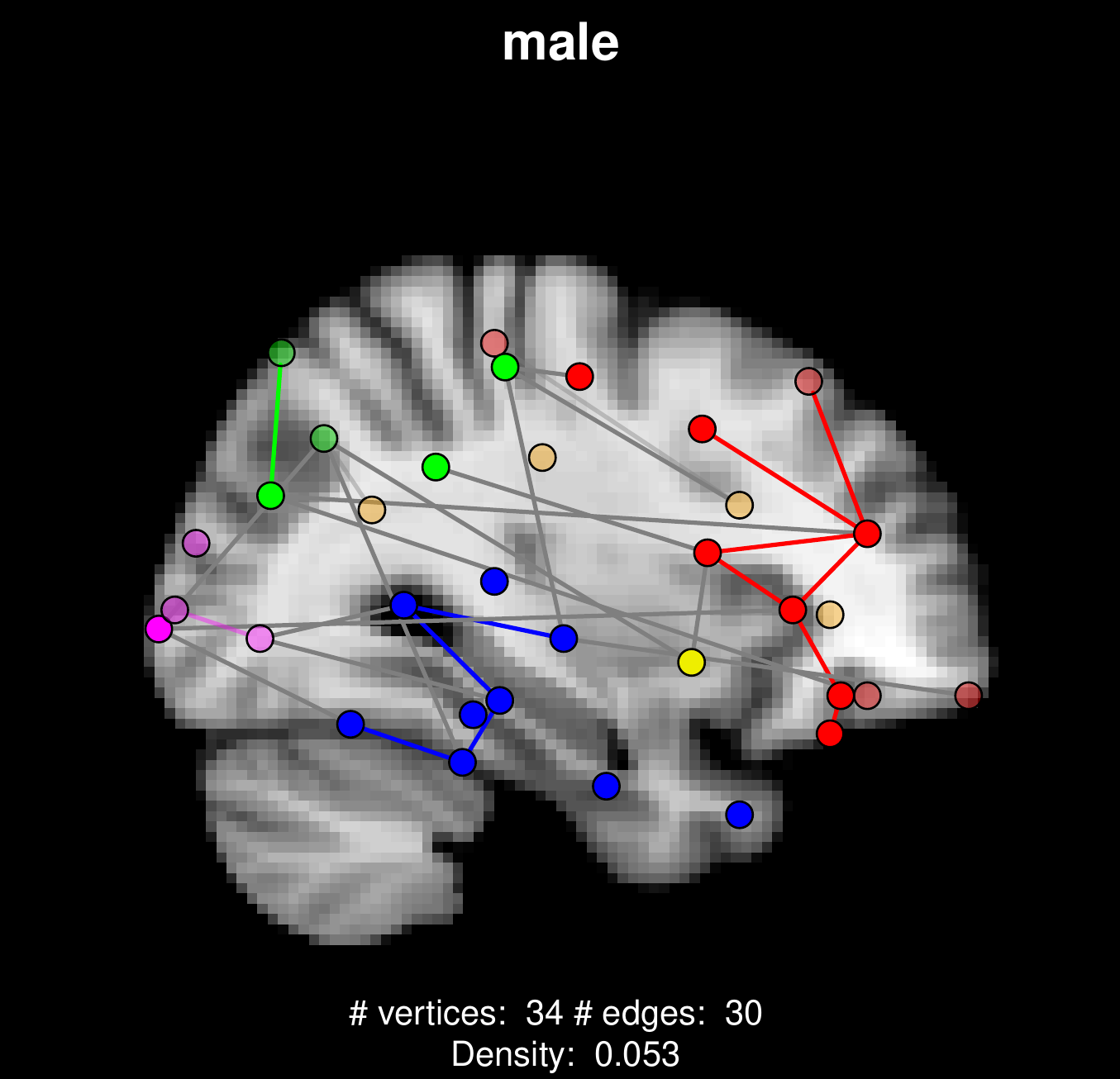}}
	\subfigure[whole]{
		\centering
		\includegraphics[width=0.3\linewidth]{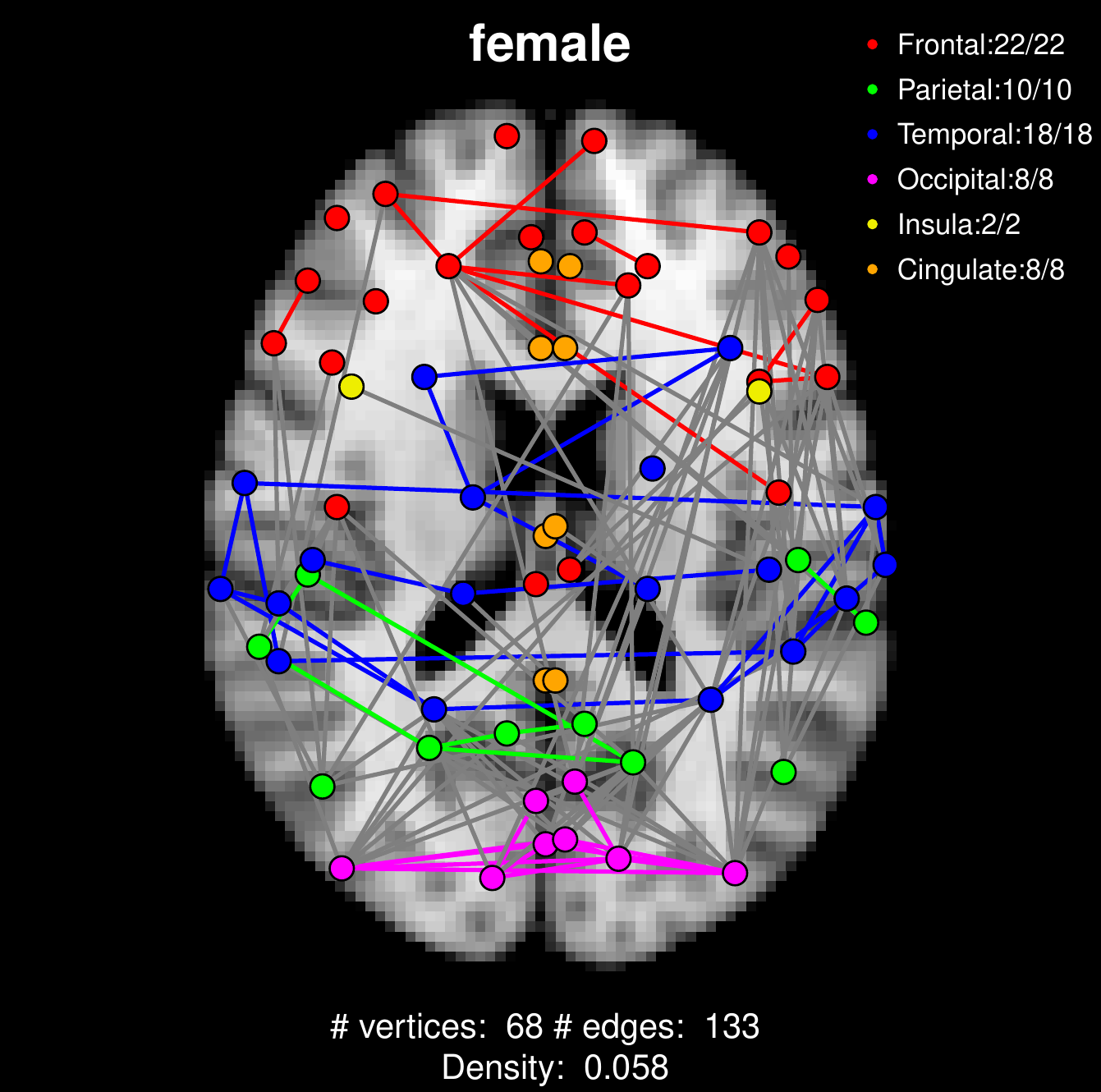}}
	\centering
		\subfigure[left]{
		\centering
		\includegraphics[width=0.31\linewidth]{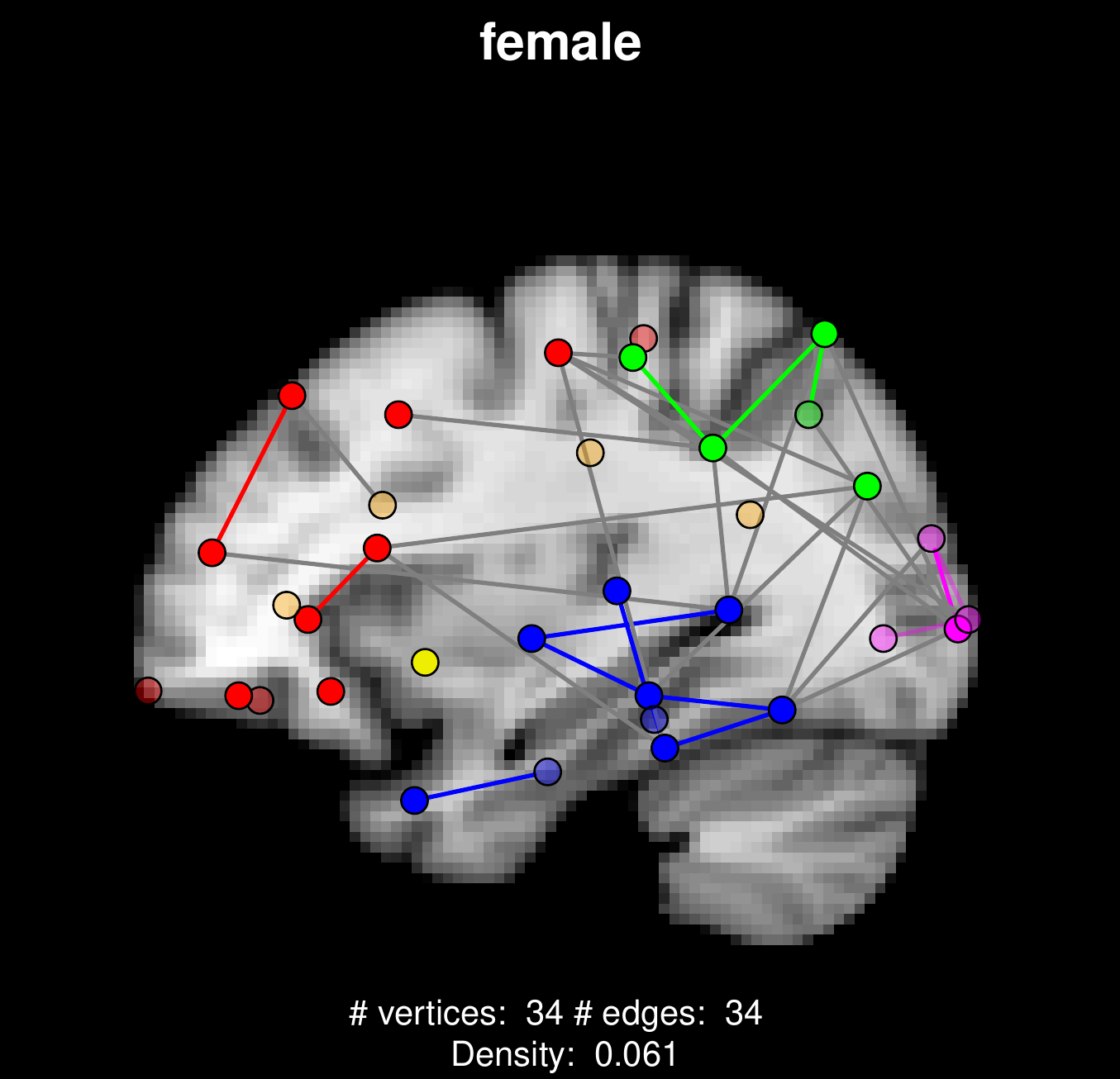}}
			\subfigure[right]{
		\centering
		\includegraphics[width=0.31\linewidth]{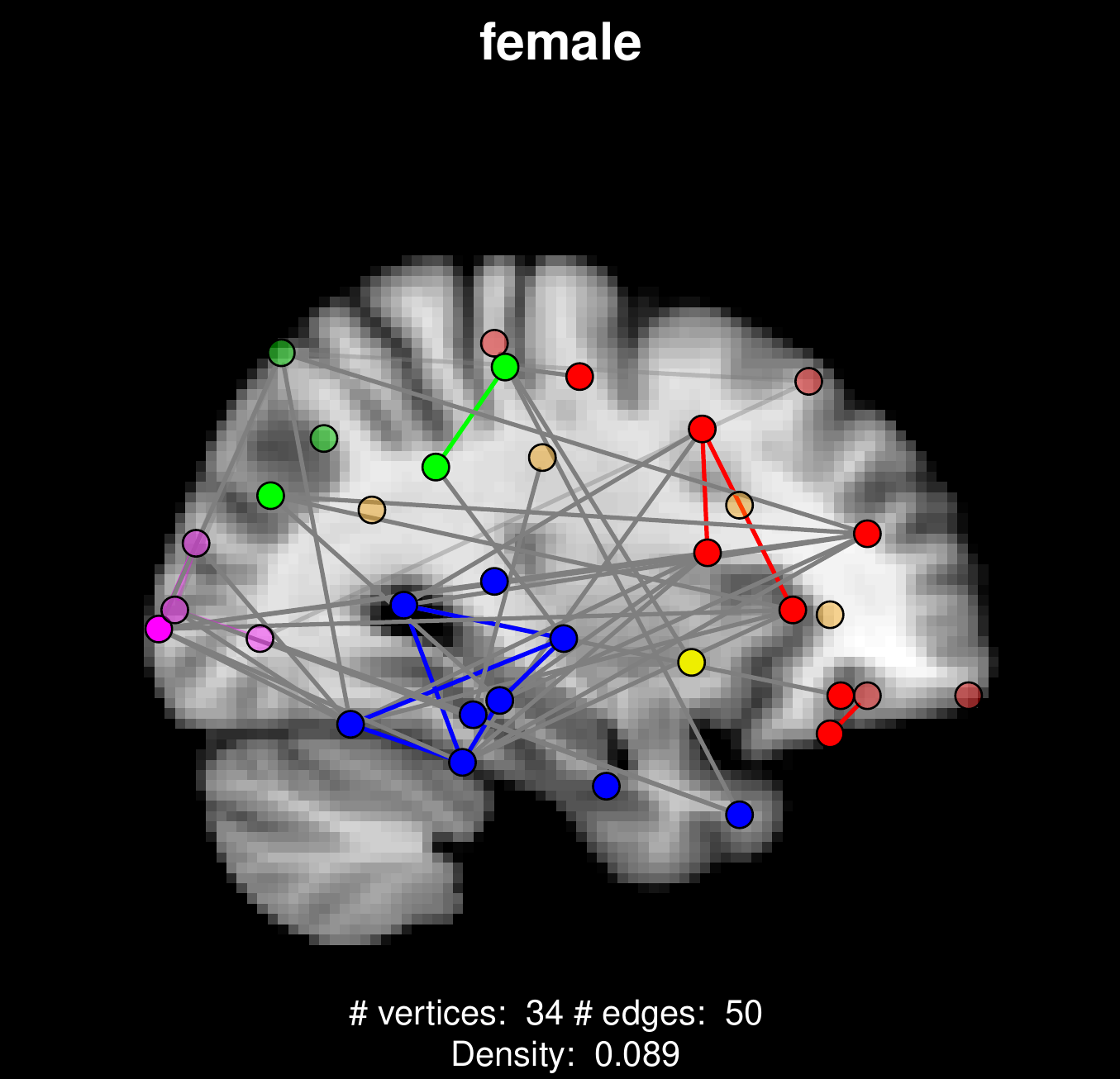}}
	\centering
	\caption{Brain network of the first frontal slice of $\hat{\mathcal{B}}_1$ based on the 6 main lobes of human brain. From the top to bottom panel, it is for male and female, respectively. Inter-lobe connections are shown in gray, and intra-lobe connections are shown in the same color as lobes. } \label{beta1}
\end{figure}

Figure \ref{beta1} visualizes the network connections of the human brain in terms of the 6 main lobes, i.e., frontal, parietal, temporal, occipital, insula and cingulate. We discover that females have greater activity in the across-lobe connectivity, particularly among the temporal, parietal, and occipital lobes \citep{ingalhalikar2014sex}, see Figures \ref{beta1} (a) and (d). 

\begin{figure}[!t]
		\centering
		\includegraphics[width=0.7\linewidth]{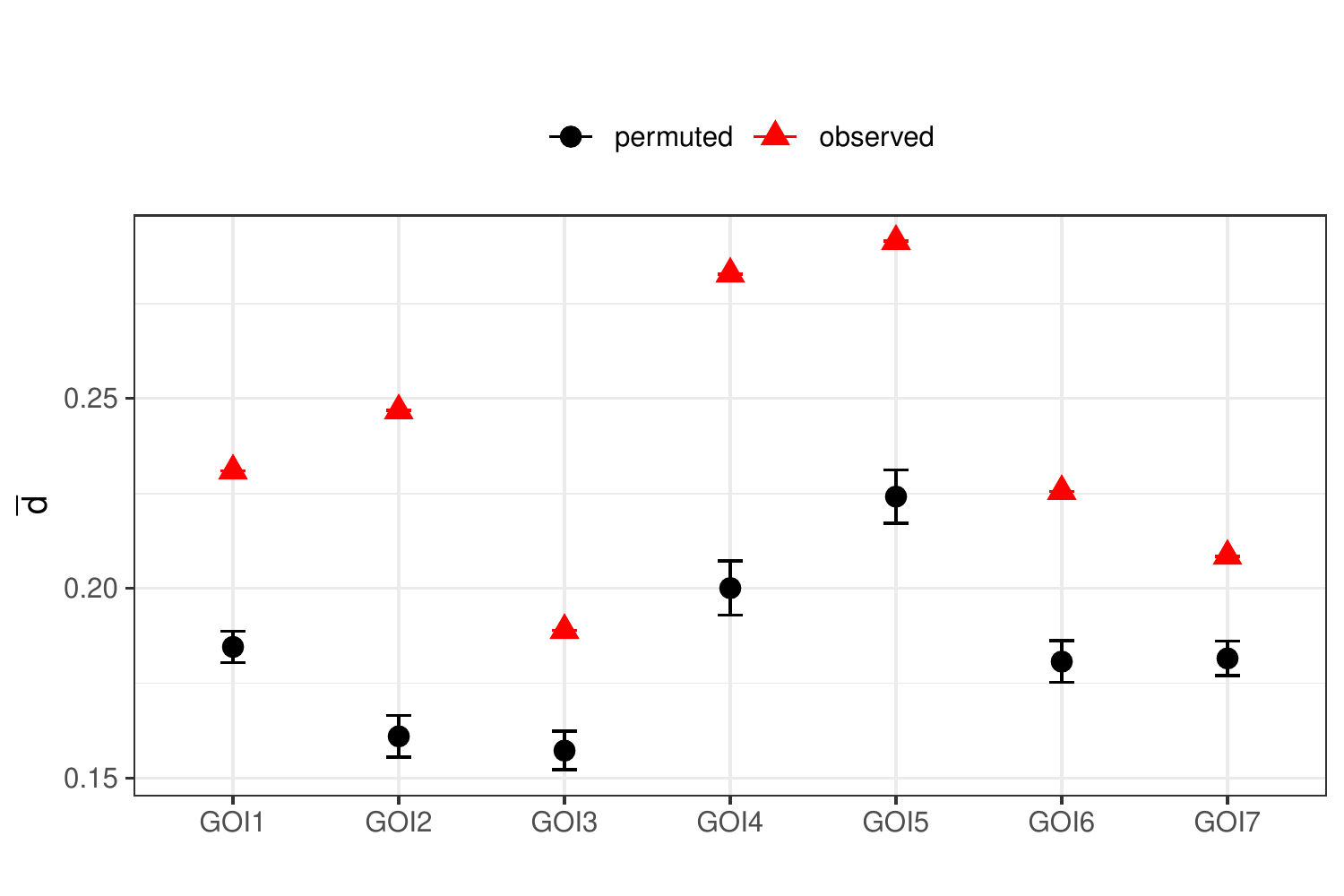}
			\centering
	\caption{The plots of $\bar{d}_{V_0}$ across 7 GOIs. Results from the permuted data are shown in black dots (with standard error bars) and the results from the observed data are shown in red triangles. 
	} \label{boxplot}
\end{figure}

In the permutation procedure, we also define $L_2$ distances for graphs of interest (GOIs). Given a set of nodes $V_0$ as 
$$\bar{d}_{V_0}=\frac{\sum_{jj' \in V_0}\D_{jj'}}{\sum_{jj' \in V_0}1(\D_{jj'}>0)},$$
where $\D$ refers to the distance $\D^{\text{obs}}$ or $\D^{\text{per},i}$. We calculate $\bar{d}_{V_0}$ based on 7 GOIs, defined as
\begin{itemize}
    \item[] GOI 1: the entire brain
    \item[] GOI 2: community 4 within the right hemisphere
    \item[] GOI 3: community 4 within the left hemisphere
    \item[] GOI 4: community 2 within the right hemisphere
    \item[] GOI 5: community 2 within the left hemisphere
    \item[] GOI 6: between community 2, right hemisphere and community 4, left hemisphere
    \item[] GOI 7: between community 2, left hemisphere and community 4, right hemisphere.
\end{itemize}
Figure \ref{boxplot} compares the $\bar{d}_{V_0}$'s calculated from the observed data and the permuted data across the above 7 GOIs. 
It is seen that the sex-specific differences from the observed data are consistently greater than those from permuted data. 

\section{The ROIs in the Desikan-Killiany atlas}
\begin{table}[t!]
	\caption{The 68 ROIs in the Desikan-Killiany atlas organized into 6 brain lobes: Temporal, Frontal, Occipital, Parietal, Cingulate, and Insula.}
	\label{68region}
	\centering
 
	\begin{tabular}{|l|p{12cm}|}
		\hline
		Temporal& 1-Left bankssts,
  5-Left entorhinal, 6-Left fusiform, 8-Left inferior temporal, 14-Left middle temporal, 15-Left parahippocampal, 29-Left superior temporal, 32-Left temporal pole, 33-Left transverse temporal, 35-Right bankssts, 
  39-Right entorhinal, 40-Right fusiform, 42-Right inferior temporal, 48-Right middle temporal, 49-Right parahippocampal, 63-Right superior temporal, 66-Right temporal pole, 67-Right transverse temporal \\ 

	\hline
	Frontal & 3-Left caudal middle frontal, 11-Left lateral orbitofrontal, 13-Left medial orbitofrontal, 16-Left paracentral, 17-Left pars opercularis, 18-Left pars orbitalis, 19-Left pars triangularis, 23-Left precentral, 26-Left rostral middle frontal, 27-Left superior frontal, 31-Left frontalpole, 37-Right caudal middle frontal, 45-Right lateral orbitofrontal, 47-Right medial orbitofrontal, 50-Right paracentral, 51-Right parsopercularis, 52-Right parsorbitalis, 53-Right parstriangularis, 57-Right precentral, 60-Right rostral middle frontal, 61-Right superior frontal, 65-Right frontalpole \\
	\hline
	Occipital& 4-Left cuneus, 10-Left lateral occipital, 12-Left lingual, 20-Left pericalcarine, 38-Right cuneus, 44-Right lateral occipital, 46-Right lingual, 54-Right pericalcarine  \\
	\hline
	Parietal&7-Left inferior parietal, 21-Left postcentral, 24-Left precuneus, 28-Left superior parietal, 30-Left supramarginal, 41-Right inferior parietal, 55-Right postcentral, 58-Right precuneus, 62-Right superior parietal, 64-Right supramarginal \\
	\hline
	Cingulate &2-Left caudal anterior cingulate, 9-Left isthmus cingulate, 22-Left posterior cingulate, 25-Left rostral anterior cingulate, 36-Right caudal anterior cingulate, 43-Right isthmus cingulate, 56-Right posterior cingulate, 59-Right rostral anterior cingulate \\
			\hline
   Insula& 34-Left insula, 68-Right insula \\
   \hline
	\end{tabular}
\end{table}

\end{document}